\documentclass[iop]{emulateapj}
\usepackage{graphicx,epsfig,amssymb,color,lineno}
\usepackage{natbib}
\usepackage{amsmath}
\usepackage{enumitem}

\slugcomment{\footnotesize Accepted for publication in ApJ}
\shorttitle{UV Luminosity Function at $1<z<3$}
\shortauthors{Alavi et al.}

\begin{document}

\title{THE EVOLUTION OF THE FAINT END OF THE UV LUMINOSITY FUNCTION DURING THE PEAK EPOCH OF STAR FORMATION ($1<\lowercase{z}<3$)\footnotemark[1]} \footnotetext[1]{Some of the data presented herein were obtained at the W.M. Keck Observatory, which is operated as a scientific partnership among the California Institute of Technology, the University of California and the National Aeronautics and Space Administration. The Observatory was made possible by the generous financial support of the W.M. Keck Foundation.}

\author{\sc Anahita Alavi\altaffilmark{2}, Brian Siana\altaffilmark{2}, Johan Richard\altaffilmark{3}, Marc Rafelski \altaffilmark{4,5}, Mathilde Jauzac \altaffilmark{6,7,8}, Marceau Limousin\altaffilmark{9}, William R. Freeman\altaffilmark{2}, Claudia Scarlata\altaffilmark{10}, Brant Robertson\altaffilmark{11}, Daniel P. Stark\altaffilmark{12}, Harry I. Teplitz\altaffilmark{12}, Vandana Desai\altaffilmark{13}}

\altaffiltext{2}{Department of Physics and Astronomy, University of California, Riverside, CA 92521, USA} 
\altaffiltext{3}{Centre de Recherche Astrophysique de Lyon, Universit\'e Lyon 1, 9 Avenue Charles Andr\'e, F-69561 Saint Genis Laval Cedex, France}
\altaffiltext{4}{Goddard Space Flight Center, Code 665, Greenbelt, MD 20771, USA}
\altaffiltext{5}{Space Telescope Science Institute, Baltimore, MD, USA}
\altaffiltext{6}{Centre for Extragalactic Astronomy, Department of Physics, Durham University, Durham DH1 3LE, U.K.}
\altaffiltext{7}{Institute for Computational Cosmology, Durham University, South Road, Durham DH1 3LE, U.K.}
\altaffiltext{8}{Astrophysics and Cosmology Research Unit, School of Mathematical Sciences, University of KwaZulu-Natal, Durban 4041, South Africa}
\altaffiltext{9}{Aix Marseille Univ, CNRS, LAM, Laboratoire d'Astrophysique de Marseille, Marseille, France}
\altaffiltext{10}{Minnesota Institute for Astrophysics, University of Minnesota, Minneapolis, MN 55455, USA}
\altaffiltext{11}{Department of Astronomy and Astrophysics, University of California, Santa Cruz, 1156 High Street, Santa Cruz, CA 95064}
\altaffiltext{12}{Department of Astronomy, Steward Observatory, University of Arizona, 933 North Cherry Avenue, Rm N204, Tucson, AZ, 85721}
\altaffiltext{13}{Infrared Processing and Analysis Center, Caltech, Pasadena, CA 91125, USA}

\begin{abstract}

We present a robust measurement of the rest-frame UV luminosity function (LF) and its evolution during the peak epoch of cosmic 
star formation at $1<z<3$. We use our deep near ultraviolet imaging from WFC3/UVIS on the {\it Hubble Space Telescope} ({\it HST}) 
and existing ACS/WFC and WFC3/IR imaging of three lensing galaxy clusters, Abell 2744 and MACSJ0717 from the Hubble Frontier Field 
survey and Abell 1689. Combining deep UV imaging and high magnification from strong gravitational lensing, we use photometric redshifts 
to identify 780 ultra-faint galaxies with $M_{UV}<-12.5$ AB mag at $1<z<3$. From these samples, we identified 5 new, faint, multiply imaged 
systems in A1689. We run a Monte Carlo simulation to estimate the completeness correction and effective volume for each cluster using the 
latest published lensing models. We compute the rest-frame UV LF and find the best-fit faint-end slopes of $\alpha=-1.56\pm0.04$, 
$\alpha=-1.72\pm0.04$ and $\alpha=-1.94\pm0.06$ at $1.0<z<1.6$, $1.6<z<2.2$ and $2.2<z<3.0$, respectively. Our results demonstrate 
that the UV LF becomes steeper from $z\sim1.3$ to $z\sim2.6$ with no sign of a turnover down to $M_{UV}=-14$ AB mag. 
We further derive the UV LFs using the Lyman break ``dropout" selection and confirm the robustness of our conclusions against different 
selection methodologies. Because the sample sizes are so large, and extend to such faint luminosities, the statistical uncertainties are quite 
small, and systematic uncertainties (due to the assumed size distribution, for example), likely dominate. 
If we restrict our analysis to galaxies and volumes above $> 50\%$ completeness in order to minimize these systematics, 
we still find that the faint-end slope is steep and getting steeper with redshift, though with slightly shallower (less negative) 
values ($\alpha=-1.55\pm0.06$, $-1.69\pm0.07$ and $-1.79\pm0.08$ for $z\sim1.3$, $1.9$ and $2.6$, respectively).
Finally, we conclude that the faint star-forming galaxies with UV magnitudes of $-18.5<M_{UV}<-12.5$ covered in this 
study, produce the majority ($55\%$-$60\%$) of the unobscured UV luminosity density at $1<z<3$.
\end{abstract} 


\section{Introduction}
\label{introduction}
The galaxy luminosity function (LF) is a fundamental tool to study the formation and evolution of galaxies as the shape of the LF is 
mainly determined by the mechanisms that regulate the star formation in galaxies \citep{ree77,whi78,ben03}. Comparing the LF with the 
underlying dark matter halo mass function reveals the importance of different modes of feedback in galaxy formation, with the active galactic 
nuclei feedback dominating the bright end and supernova and radiation-driven winds dominating the faint end \citep[e.g., ][]{dek06,som08}. 
Furthermore, the LF is a key probe to assess the contribution of galaxies with different luminosities to the total light budget at different redshifts. 

As ultraviolet (UV) light is a tracer of recent star formation in galaxies, the UV LF can help determine the total star formation rate density at 
all epochs. In addition, the UV LF is one of the few galaxy observables which is directly measurable at all epochs using current telescopes. 
Over the past 20 years, many studies have been devoted to UV LF measurements at high redshifts with $z>3$ \citep{ste99,ade00,bun04,dic04,ouc04,yan04,bec06,yos06,saw06,bou07,iwa07,mcl09,ouc09,van10,bra12,cuc12,mcl13,sch13,ate14,schmidt14,ate15a,ate15b,bou15,bow15,fin15,ish15}, intermediate redshifts with $1<z<3$ \citep{dah07,red08,hat10,oes10a,cuc12,saw12,par16} including our previous work \citep[hereafter A14]{ala14}, 
as well as low redshifts with $z<1$ \citep{arn05,bud05,wyd05,hab09,ly09,cuc12}. Taken together, these measurements suggest a rise and fall in the 
history of cosmic star formation from high redshifts to the present time with a peak sometime between $1<z<3$ \citep[and references therein]{mad14}. 
Therefore the redshift range of $1<z<3$, known as the peak epoch of cosmic star formation, is a critical time in galaxy evolution.

Many wide and shallow surveys have probed the UV LF of rarer, luminous galaxies at $1<z<3$. \citet{arn05} used the WFPC2 data in the HDF-North 
and HDF-South and measured a faint-end slope of $\alpha=-1.5\pm0.2$ for the UV LF at $z=2-3$. Later, \citet{red09} used a wide ground-based survey 
covering luminosities with $L>0.05\ L^{*}$\footnote{To be consistent with other studies, we quote these limits in terms of $L_{z=3}^{*}$, i.e. $M_{1700,AB}^{*}=-21.07$, 
from \citet{ste99}.} and measured a steep faint-end slope of $z=-1.73\pm0.07$ at $z=2.3$. Following the installation of WFC3 on the {\it Hubble Space 
Telescope} ({\it HST}), \citet{oes10a} used the wide, shallow Early Release Survey \citep[ERS; ][]{win11} and measured steep faint-end slopes ($-1.46<\alpha<-1.84$) 
for the UV LFs at $z=1.0-2.5$. However, in order to study the UV LF at fainter luminosities and accurately quantify the faint-end slope, deeper surveys were needed. 
In A14 (see next paragraph for more details), we used a very deep UV observation of the Abell 1689 (hereafter A1689) cluster obtained with the WFC3/UVIS channel  
and we extended the $z\sim2$ UV LF $100\times$ fainter than previous shallower surveys ($L\sim0.0005\ L^{*}$). We concluded that the UV LF has a steep 
faint-end slope of $\alpha=-1.74\pm0.08$ with no evidence of a turnover down to $M_{UV}=-13$. \citet{par16} recently found galaxies as faint as $L>0.002\ L^{*}$ 
utilizing the CANDELS/GOODS-South, UltraVISTA/COSMOS and HUDF data. However, their estimate of the faint-end slope $\alpha=-1.32\pm0.03$ is significantly 
shallower than others. A shortcoming of these two deep surveys is that they probe a single field (A1689 in A14 and HUDF dominating the faint luminosities in \citet{par16}), 
where the field-to-field variations affect the LF measurements. In this paper, we attempt to overcome this problem by combining deep observations of three lines of sight.

Faint star-forming galaxies play a critical role in galaxy formation and evolution, because they significantly contribute to IGM metal enrichment \citep{mad01,por05}, 
are the most plausible sources of ionizing photons during the reionization epoch \citep{kuh12,rob13} and maintain the ionizing background at $z > 3$ \citep{nes13}. 
However, these faint galaxies are inaccessible at high redshifts as they lay outside of the detection limits of current surveys. One powerful way to explore these faint 
galaxies, is to exploit the magnification of strong gravitational lensing offered by foreground massive systems and thus push the detection limits to lower luminosities. 
There have been many studies of high redshift galaxies lensed by individual galaxies \citep[e.g.,][]{pet02,sia08b,sia09,sta08, jon10,yua13, vas16}. However, galaxy clusters 
acting as gravitational lenses can magnify a large area \citep[e.g.,][]{nar84,kne11}, allowing a study of many highly magnified galaxies in a single pointing. 
In A14, combining our deep observations and magnification from strong gravitational lensing from A1689 enabled us to identify background ultra-faint galaxies.

This technique of targeting lensing galaxy clusters has been extensively used since the discovery of the first gravitationally lensed arc in the Abell 370 
cluster \citep{sou87}, and has culminated with recent large surveys of lensing clusters such as the CLASH \citep{pos12} and Hubble Frontier 
Field (HFF) \citep{lot16} surveys. The HFF program obtains very deep optical and near-infrared imaging over six lensing clusters using {\it HST}/ACS 
and {\it HST}/WFC3, respectively. These deep images enable a search for the faint galaxies as opposed to the shallow CLASH data, which 
restrict the search to bright galaxies even in the case of high  magnification. In addition, the HFF primary observations are complemented with 
data from {\it Spitzer, ALMA, Chandra, XMM}, VLA, VLT and Subaru as well as our deep {\it HST}/WFC3 UV imaging in this study. Since 
the beginning of the HFF program, many groups have studied the faint-end of the UV LF at $z> 5$ up to $z=9$ \citep{ate14,ish15,ate15a,ate15b,liv16}.

There are two primary methods of identifying high redshift galaxies, via photometric redshifts and color-color selection of the Lyman break. 
Both techniques require assumptions about stellar populations, dust reddening and star formation histories. However, each technique has its advantages. 
The photometric redshift method uses the full SED whereas the Lyman break method requires fewer filters and simpler completeness corrections. 
Some groups use the Lyman break technique \citep[e.g.,][]{hat10,bou15}, while other groups prefer photometric redshifts \citep[e.g.,][]{fin15,par16}. 
A general agreement between the UV LFs from these two methods is shown both at intermediate \citep{oes10a} and high redshift studies \citep{mcl11,mcl13,sch13}. 
One of the goals of this paper is to exploit the available multiwavelength imaging to provide a comparison between the UV LFs derived with these two selection techniques. 

In this paper, we utilize the strong gravitational lensing magnification from three foreground galaxy clusters (two from the HFF program) 
in combination with our deep WFC3/UVIS imaging to construct a robust sample of faint star-forming galaxies at $1<z<3$. The study is similar to A14, 
but spanning the entire redshift range $1<z<3$, and measuring the LF behind three clusters instead of one. 
This allows us to study the evolution of the UV LF during the peak epoch of global star formation activity (i.e., $1<z<3$), and compare with previous determinations. 
The structure of this paper is as follows. In Section \ref{sec:data}, we summarize the available observations and the data reduction for each lensing cluster. 
The catalog construction and photometric redshift measurements are described in Section \ref{sec:photometry} and \ref{sec:photoz}, respectively. 
We briefly review the lens models and the multiple image identification in Section \ref{sec:models}. We present our selection criteria 
and photometric redshift samples in Section \ref{sec:selection}. This is followed in Section \ref{sec:completeness}, where we provide 
detailed description for the completeness simulation. We then discuss the UV LF measurements for the photometric redshift samples in 
Section \ref{sec:lf_photz} and for the dropout samples in Section \ref{sec:lf-lbg}. We compare the UV LFs obtained by different selection techniques, 
evolution of the UV LF and UV luminosity density in Section \ref{sec:discussion}. Finally in Section \ref{sec:conclusion}, we provide a summary 
of our conclusions. In the appendices, we describe our color-color selection criteria, the corresponding LBG samples and the completeness 
simulation for the LBG UV LF. We also provide a list of newly found multiple images of A1689. 

In this paper, all distances and volumes are in comoving coordinates. All magnitudes are quoted in the AB system \citep{oke83} and we adopt $\Omega_{M}=0.3$, $\Omega_{\Lambda}=0.7$ and $H_{ 0}=70$ km s$^{-1}$ Mpc$^{-1}$.

\section{Data}
\label{sec:data}
In this section, we describe the data sets of three lensing fields used in this study and briefly explain the data reduction processes, 
as a more detailed description will be included in a future UV survey paper (Siana et al., in preparation). In this work we use deep {\it HST} 
imaging of three lensing clusters in a wide wavelength range, from UV to NIR, as described below.
  
\subsection{Hubble Frontier Field Observations and Data Reduction}
\label{subsec:ff_obs}

The HFF survey uses the {\it HST} Director's Discretionary time (GO/DD 13495, PI Lotz), to obtain deep WFC3/IR and ACS/WFC 
images of six lensing clusters and their parallel fields \citep{lot16}. The two HFF clusters analyzed here, Abell 2744 (hereafter A2744) 
and MACSJ0717.5+3745 (hereafter MACSJ0717), were observed during cycles 21 and 22, with 140 orbits of ACS/WFC and WFC3/IR 
imaging for each cluster/field pair. The NIR images are taken in F105W, F125W, F140W and F160W filters, and the optical data are obtained in 
F435W, F606W and F814W filters for each cluster. 

In addition, we obtained deep near ultraviolet images in F275W (8 orbits) and F336W(8 orbits) for three HFF clusters (including A2744 
and MACSJ0717) using the WFC3/UVIS channel onboard {\it HST}. These deep UV images are part of {\it HST} 
program ID 13389 (PI: B. Siana), which were taken between November 2013 and April 2014.

The Space Telescope Science Institute (STScI) handles the reduction and calibration of the optical and NIR images of the HFFs and 
releases the final mosaics in the Mikulski Archive for Space Telescopes (MAST)\footnote{\url{https://archive.stsci.edu/prepds/frontier/}}. 
We used the version 1.0 release of the public optical and NIR mosaics with a pixel scale of 60 mas pixel$^{-1}$. To make these mosaics, 
the raw optical and NIR exposures were initially calibrated using PYRAF/STSDAS \texttt{CALACS} and \texttt{CALWF3} programs, respectively. 
The calibrated images were then aligned and combined using \texttt{Tweakreg} and \texttt{AstroDrizzle} \citep{gon12} tasks in PYRAF/DrizzlePac 
package, respectively. In order to further improve the data reduction processes, the HFF team provides ``self-calibrated" 
ACS images including more accurate dark subtraction and charge transfer efficiency (CTE) correction as well as WFC3/IR images corrected for the time-variable sky lines.

To calibrate the raw UV data, we applied two major improvements in addition to the standard WFC3/UVIS calibration approach. 
The first improvement is related to the CTE degradation of the UVIS CCD detectors. This degradation caused by radiation damage in the CCDs, 
results in a loss of source flux and affects the photometry and morphology measurements especially in low background 
images \citep[e.g UV data,][]{tep13}. To correct for these charge losses in our UV images, we used a pixel-based CTE correction 
tool provided on the STScI website\footnote{\url{http://www.stsci.edu/hst/wfc3/tools/cte$\_$tools}}. The second improvement 
is in the dark current subtraction from the UV images. As shown in a recent work by \citet{tep13}, the standard WFCS3/UVIS dark 
subtraction process is not sufficient for removing dark structures and hot pixels, mainly due to the low background level in the UV data. 
This regular technique leaves a background gradient and blotchy patterns in the final science image. Therefore, we used a 
new methodology introduced by \citet{raf15} for subtracting the dark current and masking the hot pixels properly. 
A detailed description of this technique is presented in \citet{raf15}. 

After making the modified calibrated UV images, we use the PYRAF/DrizzlePac package to drizzle these images to the same 
pixel scale of 60 mas and astrometrically align with the optical and NIR data. The \texttt{AstroDrizzle} program subtracts the 
background, rejects the cosmic rays, and corrects the input images for the geometric distortion due to the non-linear mapping 
of the sky onto the detector. In addition to the science output images, \texttt{AstroDrizzle} generates an inverse variance 
map (IVM) which we used later to make the weight images and to calculate the image depths. A summary of all the images and their depths is given in Table \ref{tab:depth}.

\subsection{A1689 Observations and Data Reduction}
\label{subsec:a1689_obs}
In addition to the two HFF clusters, we observed the A1689 cluster. This cluster has been observed in three WFC3/UVIS 
bandpasses (F225W, F275W and F336W) as part of program IDs 12201 and 12931 (PI: B. Siana), taken in cycle 18 in 
December 2010 and cycle 20 in February and March 2012, respectively. The cycle 18 data (30 orbits in F275W, 4 orbits in F336W) 
were used in A14 to measure the UV LF of lensed, dwarf galaxies at $z\sim2$. In cycle 20, we added an F225W image (10 orbits) 
and deeper F336W data (14 orbits, for a total of 18 orbits) to expand our redshift range from $1<z<3$.

The data calibration and reduction are the same as explained above for the HFF UV images. These data are corrected for the CTE
 degradation and dark subtraction, as well. Moreover, A1689 is observed with ACS/WFC in 5 optical 
 bandpasses (F475W, F625W, F775W, F814W and F850LP), which were calibrated and reduced as was described in A14. 
 The A1689 images are all mapped to the same pixel scale of 40 mas pixel$^{-1}$.

\begin{deluxetable*}{ccccccccc}
\tablecolumns{12} 
\tablecaption{Observations and Image Depths}
\tablewidth{0pt}
\tablehead{
\colhead{Cluster} & \multicolumn{2}{c}{A2744(HFF)} & \colhead{} &  \multicolumn{2}{c}{MACSJ0717(HFF)} & \colhead{} & \multicolumn{2}{c}{A1689} \\
\hline\\
\colhead{Instrument/Filter}  & \colhead{Orbits} & \colhead{Depth\tablenotemark{a}} & \colhead{} & \colhead{Orbits} & \colhead{Depth\tablenotemark{a}} & \colhead{} & \colhead{Orbits} & \colhead{Depth\tablenotemark{a}}}
\startdata
WFC3/F225W &\nodata &\nodata &
& \nodata &\nodata &
& 10 & 27.71\\
WFC3/F275W & 8 & 27.80 &
&  8 & 27.43 &
& 30 & 28.14\\
WFC3/F336W & 8 & 28.20 & 
& 8 & 27.86& 
&18 & 28.36\\
ACS/F435W    & 18 & 28.70 &
& 19 & 28.46 &
& \nodata &\nodata \\
ACS/F475W   & \nodata &\nodata &
 &\nodata&\nodata &
 & 4 & 28.04\\
ACS/F606W    & 9 & 28.70 &
& 11 & 28.59 &
& \nodata &\nodata\\
ACS/F625W    &\nodata &\nodata &
&\nodata &\nodata & 
& 4  & 27.76\\
ACS/F775W    &\nodata &\nodata &
& \nodata&\nodata  & 
& 5 & 27.69\\
ACS/F814W    &41& 29.02 & 
& 46 & 28.87 & 
& 28 & 28.72\\
ACS/F805LP   &\nodata &\nodata &
& \nodata &\nodata & 
& 7 & 27.30  \\
 WFC3/F105W  & 24.5 & 28.97 &
 & 27 & 29.02 &
 & \nodata &\nodata \\
WFC3/F125W  & 12 & 28.64 & 
& 13 & 28.60 & 
& \nodata &\nodata \\
WFC3/F140W  & 10 & 28.76 &
& 12 & 28.61 &
& \nodata &\nodata \\ 
WFC3/F160W  & 24.5 & 28.77 & 
& 26 & 28.65 & 
& \nodata &\nodata \\
\enddata
\tablenotetext{a}{5$\sigma$ limit in a 0.2$\arcsec$ radius aperture}
\label{tab:depth}
\end{deluxetable*}

\section{Object Photometry}
\label{sec:photometry}
A detailed description for the A1689 photometry is given in A14. Here we provide the details of the photometric measurements for the HFF data.
Since our HFF data cover a large range of wavelengths (from UV up to NIR), the width of the point spread function (PSF) changes considerably.
 To do multiband photometry, we match the PSF of all of the images to the F160W band, which has the largest PSF. 
 We used the IDL routine \texttt{StarFinder} \citep{dio00} to stack all of the unsaturated stars in the field and extract the PSF. 
 We fit a simple Gaussian function to each extracted PSF using the IRAF imexamine task, and then derive the PSF matched images 
 by convolving each band with a Gaussian kernel of appropriate width. We use \texttt{SExtractor} \citep{ber96} 
 to perform object detection and photometry. The final catalog areas are 4.81, 5.74, and 6.42 arcmin$^{2}$ where the WFC3 and ACS images are available for A2744, MACSJ0717 and A1689, respectively.

 We run \texttt{SExtractor} in dual image mode, with F475W and F435W bands as detection images for A1689 and HFF 
 clusters, respectively. We use F435W band to minimize contamination from the cluster galaxies and intracluster light, 
 as this filter probes below the 4000 \AA\ break, where the galaxies are considerably fainter. To improve the 
 detection of faint objects and to avoid detecting spurious sources (i.e., over-blended from very bright galaxies), 
 for the \texttt{SExtractor} parameters, we set \texttt{DETECT$\textunderscore$MINAREA} to 4(5) 
 and \texttt{DETECT$\textunderscore$THRESH} to 0.9$\sigma$ (1.0$\sigma$) significance for A2744 (MACSJ0717). 
 The minimum contrast parameter for deblending (\texttt{DEBLENS$\textunderscore$MINCONT}) is set to 0.02 for both cluster fields. 
 The fluxes are measured in isophotal (ISO) apertures. The IVM images produced by the drizzling process as 
 mentioned in Section \ref{subsec:ff_obs}, were converted to the RMS$\textunderscore$MAPs by taking their inverse square root. 
 \texttt{SExtractor} uses these RMS$\textunderscore$MAPs to derive the flux uncertainties. We correct these 
 RMS$\textunderscore$MAPs for the correlated noise \citep[A14, ][]{cas00} from drizzling the mosaics. Finally, we correct 
 our photometry for the Galactic extinction toward each cluster using the \citet{sch11} IR dust maps. 
 To account for systematic error (i.e., due to uncertainty in the Galactic extinction, the zero point values, PSF-matched photometry), 
 we add, in quadrature, a 3$\%$ flux error \citep {dah10,var14} in all bands for all three cluster fields.

\section{Photometric Redshifts}
\label{sec:photoz}
We use a template fitting function code, \texttt{EAZY}, \citep{bram08} to estimate the photometric redshift of galaxies in all of our lensing fields. 
\texttt{EAZY} has two characteristic features that distinguish it from the other photometric redshift codes. 
First, it derives the optimized default template set from semianalytical models with perfect completeness down to very 
faint magnitudes rather than using biased spectroscopic samples. Second, it has the ability to fit to a linear combination of 
basis templates rather than fitting to a single template, which is usually not a good representation of a real galaxy. 
We varied several \texttt{EAZY} input parameters to find the optimal values. Running \texttt{EAZY} using a variety 
of empirical \citep{col80,kin96} or stellar synthetic templates \citep{gra06,bla07} allows us to find the set of models where 
the output photometric redshifts are in the best agreement with the spectroscopic redshifts. We use P\'{E}GASE \citep{fio97} 
stellar synthetic templates, which provide a self-consistent treatment of nebular emission lines and include a wide variety of 
star formation histories (constant, exponentially declining) and a Calzetti dust attenuation curve \citep{cal00}. 
We do not use template error function capability in EAZY because it causes poorer agreement with spectroscopic redshifts. 
We also do not use the magnitude priors, as these functions do not cover the faint luminosities targeted in this work. \texttt{EAZY} 
uses the \citet{mad95} prescription for absorption from the intergalactic medium. 

\begin{figure}
\epsscale{2.0}
\includegraphics[trim=0.2cm 0.1cm 0.1cm 0cm,clip=true,width=\columnwidth]{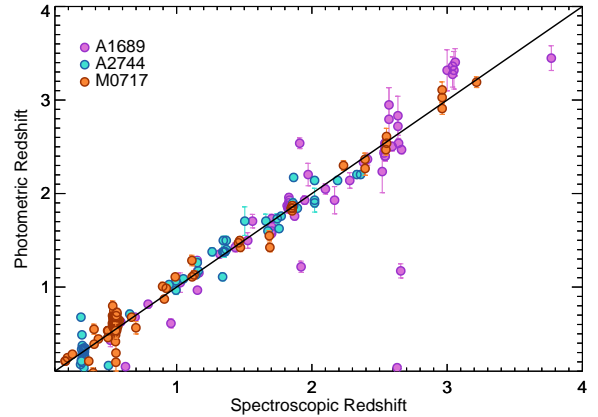}
\caption{ Comparison between the photometric and spectroscopic redshifts for 244 galaxies in all three lensing fields. 
The spectroscopic redshifts are either from our Keck/MOSFIRE and Keck/LRIS data or from the literature 
(for more details see the text). The purple, orange and cyan circles show the measurements for the A1689, MACSJ0717 and A2744 cluster fields, respectively.}
\label{fig:specz_photz}
\end{figure}

For the HFFs (A1689), we derive the photometric redshifts using the complete 9(8) photometry bands 
of F275W, F336W,F435W, F606W,F814W, F105W, F125W, F140W and F160W (F225W, F275W, F336W,F475W, F625W, F775W, F814W, F850LP) 
with the central wavelengths covering from 0.27-1.54 (0.24-0.91) $\micron$. Figure \ref{fig:specz_photz} 
shows the comparison between the photometric redshifts and the spectroscopic redshifts for all three clusters. 
For both of the HFF clusters, we use the spectroscopic redshifts from the GLASS program, which obtained grism 
spectroscopy of 10 massive clusters including the HFFs \citep{sch14,tre15}. We note that we only include their measurements 
with high quality parameter (i.e., quality$>4$) for a secure redshift estimate. In addition, for A2744, we also use the spectroscopic 
redshifts from the literature \citep{owe12,ric14,wan15}. For MACSJ0717, we add the spectroscopic redshifts from 
our Keck/MOSFIRE spectral observations as well as the redshfits from the literature \citep{lim12,eb14}. 
Most of the spectroscopic redshifts of A1689 were described in A14, but here we also include our new measurements 
from our Keck/MOSFIRE spectra taken on January 2015. A detailed study of spectroscopic 
data for these samples will be presented in a future paper. From all 186 galaxies with spectroscopic redshifts, 68 are within 
our target redshift range of $1<z<3$. For these galaxies with spectroscopic redshift of $1<z_{spec}<3$, 
we calculate normalized median absolute deviation\footnote{The normalized median absolute 
deviation is defined as $\sigma_{NMAD}=1.48\times \mathrm{median}(|\Delta z-\mathrm{median}(\Delta z)|/(1+z))$ \citep{ilb06,bram08}. 
Unlike the usual standard deviation, $\sigma_{NMAD}$ is not sensitive to the presence of outliers.} 
to be $\sigma_{NMAD}=0.025$ \citep{ilb06} and find six outliers defined to have $\Delta z/(1+z_{spec}) > 5\sigma_{NMAD}$ \citep{bram08}. 
The median and mean values of fractional redshift error, $\Delta z/(1+z_{spec})$, after excluding outliers are 0.02 and 0.03, respectively. 

Though the agreement between the photometric and spectroscopic redshifts is strong 
evidence for reliability of our redshift estimates, it is restricted to the brighter galaxies. 
While our photometric redshift samples contain galaxies as faint as F606W (F625W for A1689) $=30$ AB magnitudes, our 
spectroscopic samples cover magnitudes down to F606W (F625W for A1689) $=26.46$. 
We note that among these objects, we have 5 galaxies at $1.2<z_{spec}<2.2$ with very faint magnitudes of 
$-15.4<M_{UV}<-14$, where their spectroscopic and photometric redshifts agree well with mean 
$\Delta z/(1+z_{spec})=0.04$. To further investigate the reliability of our 
photometric redshift estimates of the faint galaxies\footnote{We define the faint 
galaxies based on the limiting magnitude used in our sample selection criteria (see Section \ref{sec:selection}). 
They are defined to have S/N$<5$ in either detection filter or the rest-frame 1500 \AA\ filter.}, where the spectroscopic 
redshifts are not available, we use a redshift quality parameter, Q $\footnote{Q parameter (see Equation 8 in \citet{bram08}) 
combines the reduced-$\chi^{2}$ of the fitting procedure with the width of the $68\%$ confidence interval of the 
redshift probability distribution function to present an estimate of the reliability of the output redshift.}$. 
It is a statistical estimate of the reliability of the photometric redshift outputs of EAZY. \citet{bram08} find that the 
photometric redshift scatter (i.e., difference between photometric redshift and spectroscopic redshift) is 
an increasing function of Q parameter with a sharp increase above Q=2-3. We calculate the Q parameter for our faint 
galaxies, as well as for the galaxies with spectroscopic redshifts of $1<z<3$. A comparison between these 
two sub-samples shows that the distributions of Q values are similar (i.e., the faint galaxies are {\it not} skewed toward higher values of Q), 
such that the spectroscopic galaxies have median Q of 0.9, 1.1 and 1.5 relative to the 
faint galaxies with median Q of 0.6, 1.1 and 2.2 for $z\sim1.3$, 2.2 and 2.6 samples, respectively. 
We note that these values are within the safe regime for Q parameter (i.e., $Q<3$, as explained above).

\section{Lens models}
\label{sec:models}
In order to estimate intrinsic properties (i.e., luminosity) of the background lensed galaxies in our samples, 
we require an accurate mass model of the galaxy cluster to calculate the lensing magnification. 
For the HFF program, there are several groups working independently to use deep HFF optical and NIR imaging to model 
the mass distribution for all of the six clusters \citep{Bra05,lie06,die07,jul07,jul09,mer09,zit09,ogu10,mer11,zit13,sen14}. 
The main distinction between these models is that some groups assume light traces mass and parametrize the total mass 
distribution as a combination of individual cluster members and large scale cluster halo components, while the other groups 
use a non-parametric mass modeling technique, avoiding any priors on the light distribution. In a recent study, \citet{pri16} 
provide a comparison between these different lens models. All of these models are constrained by the location and the redshift 
of known multiply imaged systems. Besides observational constraints from strong gravitational lensing, several teams also 
incorporate the weak lensing shear profile from ground-based observations. All of the HFF lens models and the methodologies 
adopted by each team are publicly available via the STScI website\footnote{https://archive.stsci.edu/prepds/frontier/lensmodels/}. 
In this section, we briefly review the mass models that we used for each of our lensing clusters.

\subsection{HFF Lensing Models}
\label{subsec:HFF lensing models}
For the HFF clusters, we utilize the lens models produced by the Clusters As TelescopeS (CATS) 
collaboration (Co-PIs J.-P. Kneib and P. Natarajan; Admin PI H. Ebeling) who use the \texttt{Lenstool} 
software\footnote{https://projets.lam.fr/projects/lenstool/wiki} \citep{jul07} to parameterize the lens mass distribution. 
\texttt{Lenstool} is a hybrid code which combines both strong- and weak-lensing data to constrain the lens mass model. 
\texttt{Lenstool} models each cluster's mass as a composition of one or more large cluster halos plus smaller 
subhalos associated with individual galaxies identified either spectroscopically or photometrically as cluster members. 
The output best model from \texttt{Lenstool} is parameterized through a Bayesian approach. 

For A2744, we use the strong lensing model of \citet{jau15}, which uses 61 multiply-imaged systems found in 
the complete HFF optical and NIR data. For MACS J0717, we use the strong lensing model of \citet{lim16}, which uses 55 
multiply-imaged systems found in the complete HFF optical and NIR data.

\subsection{A1689 Lensing Model}
\label{subsec:A1689 lensing models}
As in A14, the lens model that we use for A1689 is from \citet{lim07}. Similar to the mass reconstruction 
techniques for the HFF clusters, \citet{lim07} optimize a parametric model implemented in the \texttt{Lenstool} 
using 32 multiply imaged systems behind A1689. Their optimized lens model for A1689 is a composite of two large-scale halos and the subhalos of cluster member galaxies.

\subsection{Multiply imaged systems}     
\label{subsec:multi}

Finding more multiply imaged systems is critical for improving a lens model, as the lens model is constrained by 
the location and redshift of these systems. In addition, identifying the multiple images is important as we need to 
remove them from the galaxy number counts. We run \texttt{Lenstool} using each previously described lens model as 
an input, to look for the potential counter-images for each lensed galaxy in the sample.

 Currently, there is no automated process for identifying multiple imaged systems. Here, we summarize the 
 approach that we took to find new multiply imaged systems. 1) We run \texttt{Lenstool} entering the coordinates 
 and the photometric redshift of each galaxy to predict the location of its potential counter-images. 
 In this step, \texttt{Lenstool} first de-lenses the galaxy image to its original position in the source plane at the 
 given photometric redshift, and then re-lenses it back to all of the possible multiple image positions in the 
 image plane. 2) We search for the objects with the same color and symmetry in the morphology near 
 the predicted positions. 3) If we find any nearby candidate from step 2, we then repeat the first step to check if 
 the potential counter-images of the candidate match with the first object. 4) Finally, we require the same 
 photometric redshifts (within 1$\sigma$ accuracy), for all of the newly found multiple images. 
 This final criterion exhibits the importance of covering rest-frame UV wavelengths, which enables us to identify 
 the Lyman break to distinguish the high redshift objects (in this case $1<z<3$) from the lower redshift interlopers, 
 since both often have flat, featureless SEDs at rest-frame optical wavelengths.

Following this procedure for all the galaxies, we find 5 and 3 new multiply imaged systems behind A1689 and MACSJ0717, 
respectively. 
Our new findings in cluster MACSJ0717 added new systems 21, 80 and 82 to the list reported in \citet{lim16}. 
We introduce the new A1689 multiply imaged systems in appendix \ref{appendixC}.

\section{Sample selection}
\label{sec:selection}
We use the photometric redshift estimates to construct our galaxy samples in three redshift ranges 
of $1.0<z<1.6$, $1.6<z<2.2$ and $2.2<z<3.0$. To ensure the reliability of our photometric redshifts 
and to avoid selecting spurious objects in the sample, we require 3$\sigma$ detections in the detection 
filter and the rest-frame 1500 \AA\ filter. The selection criteria for the lower redshift range are:

\begin{enumerate}[itemsep=0mm,label=\alph*.]
\item$1.0<z_{phot}<1.6$
\item$S/N>3$ in the F275W and F336W bands.
\end{enumerate}
 selecting 70, 134 and 93 candidates in A1689, A2744 and MACSJ0717, respectively. 
 The selection criteria for the middle redshift range are:

\begin{enumerate}[itemsep=0mm,label=\alph*.]
\item$1.6<z_{phot}<2.2$
\item$S/N>3$ in the F336W and F435W (F475W) bands for the HFFs (A1689).
\end{enumerate}
selecting 128, 121 and 69 candidates in A1689, A2744 and MACSJ0717, respectively. And finally, 
the selection criteria for the higher redshift range are:

\begin{enumerate}[itemsep=0mm,label=\alph*.]
\item$2.2<z_{phot}<3.0$
\item$S/N>3$ in the F435W and F606W bands for the HFFs.
\end{enumerate}

selecting 176 and 102 galaxies in the A2744 and MACSJ0717 fields, respectively. We should note that we do 
not include data from A1689 for the highest redshift ($z\sim2.6$) analysis because, due to the cluster redshift 
of $z=0.18$, the Balmer break of faint cluster members \citep[like globular clusters, ][]{alam13} can mimic the Lyman break at $z\sim3$.

 In total, we have 297, 318 and 278 candidates at $1<z<1.6$, $1.6<z<2.2$ and $2.2<z<3.0$, respectively. 
 As explained in Section \ref{subsec:multi}, we must clean our samples of multiple images. Among each multiply
  imaged system, we keep the brightest image and remove the rest of the images from our samples. However,
   if the brightest image has a magnification higher than 3.0 magnitudes, we then select the next brightest image. 
   This condition on magnification is considered to ensure the reliability of the magnification value predicted from the lensing models. 

Furthermore, to ensure purity of the samples, we consider different possibilities of contamination in the photometric 
redshift selected samples. First, to find possible contamination from stars, we use the \citet{pic98} stellar spectra 
library to predict stellar colors for a variety of stars and compare with the color of our candidate galaxies. 
In the case of similar colors, we visually inspect the objects. We found only 1 ($\sim0.3\%$), 2 ($\sim0.6\%$) 
and 0 stars in the $1.0<z<1.6$, $1.6<z<2.2$ and $2.2<z<3.0$ samples, respectively. We also visually inspect 
all of the galaxies to exclude objects associated with diffraction spikes and nearby bright galaxies. 
The contamination is only 2 ($\sim0.7\%$), 3 ($\sim0.9\%$) and 0 for the $1.0<z<1.6$, $1.6<z<2.2$ and $2.2<z<3.0$ 
samples, respectively. Finally, after excluding all of the multiple images and the contamination, we have 277, 269 
and 252 galaxies at $1.0<z<1.6$, $1.6<z<2.2$ and $2.2<z<3.0$, respectively.

With the aim to measure the UV luminosity function, we use the F336W, F435W and F606W bands for the HFFs and F336W 
and F475W bands for the A1689 samples to measure the absolute magnitude at rest-frame 1500 \AA\ ($M_{UV}=M_{1500}$) 
at redshifts $1.0<z<1.6$, $1.6<z<2.2$ and $2.2<z<3.0$, respectively. As we did in A14, we determine the intrinsic 
absolute magnitudes of $M_{1500}$ by applying the magnification corrections computed from the lens models discussed in Section \ref{sec:models}.
\begin{equation}
M_{1500}=m+\mu_{\mathrm{mag}}-5\text{log}(d_{L}/10\text{ pc})+2.5\text{log}(1+z)
\label{eq:M}
\end{equation}

\begin{figure}
\epsscale{2.0}
\includegraphics[trim=0cm 0cm 0cm 0cm,clip=true,width=\columnwidth]{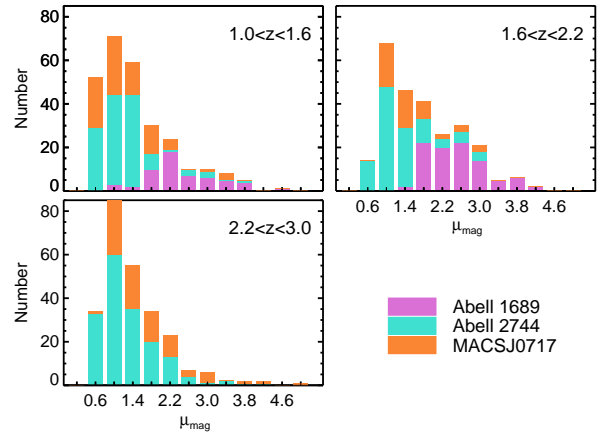}
\caption{The magnification distribution of galaxies expressed in magnitude units. The purple, orange and cyan colors show 
the number of candidate galaxies for each magnification bin on A1689, A2744 and MACSJ0717, respectively. 
These clusters provide a large range of magnifications, 
with higher values mostly from A1689 (see the text).}
\label{fig:hist_magn}
\end{figure}

\begin{figure}
\epsscale{2.0}
\includegraphics[trim=0cm 0cm 0cm 0cm,clip=true,width=\columnwidth]{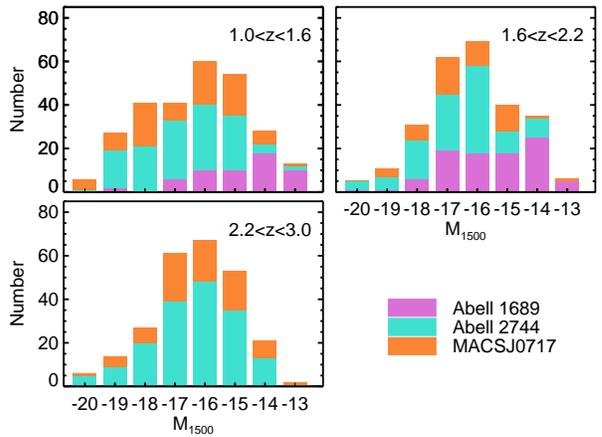}
\caption{The intrinsic absolute UV magnitude (i.e., corrected for the lensing magnification) distribution of all galaxies 
in our three redshift slices. The colors are similar to Figure \ref{fig:hist_magn} and they show the number of candidate galaxies for each absolute magnitude 
bin. We cover a large luminosity range by combing the HFFs with A1689, 
which finds the faintest galaxies (${M_{1500} > -14.5}$).}
\label{fig:hist_abs_mag_new}
\end{figure}

Where $\mu_{\mathrm{mag}}$ is the predicted magnification in magnitude units from the lensing model of each cluster. 
We limit our samples to galaxies brighter than $M_{1500} < -12.5$ magnitudes, to ensure a reliable absolute magnitude 
measurement. All of the galaxies brighter than this limit have magnification uncertainty from lensing models below 0.5 
magnitudes with mean value of 0.03 magnitudes. But the galaxies fainter than $M_{1500} > -12.5$ have magnification 
uncertainties above 2.0 magnitudes. This limit excludes 7 ($2.5\%$), 10 ($3.7\%$) and 1 ($0.4\%$) galaxies from 
the $1.0<z<1.6$, $1.6<z<2.2$ and $2.2<z<3.0$ samples, respectively. 

Figure \ref{fig:hist_magn} shows the distributions 
of magnifications for galaxies in three photometric redshift samples. The magnification values range 
between $\mu_{\mathrm{mag}} =0.5-4.8$ (equivalent to $1.58-83$ in flux density units) with median values
 of $\mu_{\mathrm{mag}}=1.27$, 1.61 and 1.24 for $z\sim1.3$, 1.9 and 2.6 samples, respectively. 
 As shown in Figure \ref{fig:hist_magn}, most of the highly magnified galaxies ($\mu_{\mathrm{mag}}>2.5$) in the $z\sim1.3$ 
 and $z\sim1.9$ samples are from A1689. We note that, because A1689 has a large Einstein radius, 
 it provides high magnification (i.e., median $\mu_{\mathrm{mag}}=2.0$) over large area in the source plane. 
 Therefore, objects with high magnification in A1689 are not required to 
 be close to the critical lines, where the magnification formally diverges. For example, the galaxies with high magnification 
 ($\mu_{\mathrm{mag}}=2.5-4$) in the A1689 sample are on average 15 arcsec (with median of 10 arcsec) from the critical lines whose 
 positions are predicted with a precision of 2.87 arcsec by the lens model \citep{lim07}. Therefore, these magnification estimates are not strongly 
 affected by uncertainties in the location of the critical lines.

In Figure \ref{fig:hist_abs_mag_new}, we show the histograms of absolute UV magnitudes for each lensing 
cluster in three redshift bins. This figure emphasizes the importance of including A1689, since it dominates the 
number of galaxies at the faintest magnitudes, $M_{1500}>-14.5$.

\section{Completeness Simulations}
\label{sec:completeness}
In order to connect the observed galaxies to the underlying population of all star-forming galaxies, we need to 
precisely estimate the completeness of our sample. This is more critical for low luminosity bins, where the galaxies are close to the detection limits. 
An approach commonly used in the blank field studies to estimate the completeness \citep[e.g.,][]{oes10a,gra11,bow15, fin15}, 
is to generate artificial galaxies with properties similar to the real galaxies and then apply an identical selection technique as 
for the observed candidates to calculate the fraction of recovered simulated galaxies in a given magnitude and redshift bin. 
This technique is also applicable in gravitationally lensed studies \citep[e.g.,][]{ate15a,ate15b}. However, one needs to 
incorporate the added complexity due to the strong lensing amplification. 

In this work, we adopt a Monte Carlo simulation following the methodology presented in detail in A14. Here, we briefly 
describe these completeness simulations, and we provide additional details where our approach deviates from what was done in A14.
 
We compute the completeness in a 3-D grid of redshift, magnitude and lensing magnification. For each point in this 3-D space, 
we assign a redshifted and magnified template galaxy spectrum, which is generated by \citet{bru03} (hearafter BC03) synthetic 
stellar population models assuming a $0.2\ Z_{\odot}$ metallicity and an age of 100 Myr. A detailed justification for these 
assumptions is given in A14. The SED is dust attenuated using the Calzetti extinction curve \citep{cal00} and a random 
color excess, E(B-V), value taken from a Gaussian distribution centered at 0.15 as measured in A14 and other 
studies \citep{ste99, red09, hat13} with a standard deviation of 0.1. In order to understand the effect of a changing 
reddening distribution, we also examined the completeness for a model in which the dust reddening linearly decreases 
toward fainter luminosities. To derive this linear function, we measured the relation between UV spectral slope 
and $M_{1500}$ magnitude for our galaxies and we calculate the dust reddening values assuming a Calzetti reddening 
curve. The final completeness corrections from this examination show only negligible changes relative to our original simulations.
 \footnote{We note that for the same experiment, the effective volumes of the LBG samples (see Appendix \ref{lbg_comp}) show slightly larger change at bright 
 luminosities. This can be understood by considering that the color-color criteria select against very reddened galaxies. 
 However, our final estimates of the best-fit LFs (for both sample selections) are robust against these different initial assumptions of dust reddening distribution.}

We then create transmission curves (as a function of wavelength) for 300 lines of sight through the intergalactic 
medium (IGM) at that redshift. The IGM opacity is calculated using a Monte Carlo simulation to randomly place Hydrogen 
absorbers in each line of sight as described in A14 \citep[see also][]{sia08a}. Our completeness simulation is modified relative to A14 in the following two ways.\\
  
\begin{figure}
\epsscale{2.0}
\includegraphics[trim=1cm 1cm 3cm 3cm,clip=true,width=1.\columnwidth]{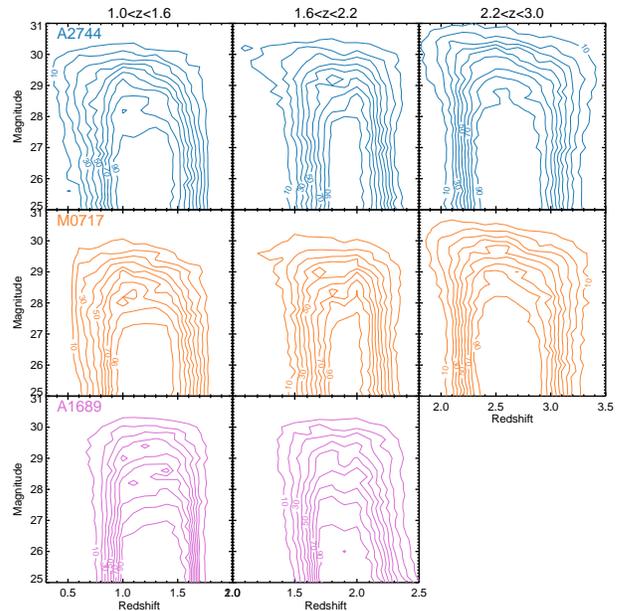}
\caption{Completeness as a function of intrinsic apparent magnitude on the y-axis and redshift on the x-axis. 
The blue, orange and purple contours show the completeness simulation values for A2744, MACSJ0717 and A1689, 
respectively. The left, middle and right columns represent the completeness contours for photometric redshift 
samples at $1.0<z<1.6$, $1.6<z<2.2$ and $2.2<z<3.0$, respectively.  These contours are drawn for magnification of $\mu_{\mathrm{mag}}=2.0$ magnitudes.}
\label{fig:Completeness}
\end{figure}
  
{\it Updating the Size Distribution of Star-forming Galaxies}: One of the key factors in estimating the incompleteness 
is the assumed size distribution for galaxies. As shown in \citet{gra11}, the completeness correction at low luminosities 
depends critically on the adopted size distribution in the simulation, as using too small (large) a size distribution can 
cause one to over- (under-) estimates the completeness. As reported in various observational 
studies \citep[e.g.,][]{bou04,fer04,hua13}, the rest-frame UV sizes of high redshift Lyman break galaxies follow a log-normal 
distribution. In a recent work, \citet{shi15} measured the size distribution of a large sample of galaxies at $0<z<8$, 
using the 3D-HST and CANDELS data. They showed that the circularized effective radius\footnote{The circularized 
effective radius is defined as $r_{e}=r_{e,major}\sqrt{q}$, where $r_{e,major}$ is the half-light radius along the semi-major 
axis and q is the axis ratio. The circularized radius has been extensively used in other high redshift size 
measurements \citep[e.g.,][]{mos12, ono13}} ($r_{e}$) distribution of star-forming galaxies at $0<z<8$ is well represented by 
a log-normal distribution whose median decreases toward high redshifts (at a given luminosity) and changes with 
luminosity as $r_{e}\propto (L_{UV})^{\alpha}$ with $\alpha=0.27$ for all redshifts. For our completeness simulation, 
we generate random galaxy sizes at each luminosity and redshift using the corresponding log-normal distribution from 
Table 8 in \citet{shi15}. We extrapolate their measurements below $M_{UV}<-16$. Using the randomly selected $r_{e}$ values, 
we then adopt a Sersic profile with index n=1.5 as suggested by \citet{shi15}. Other LF studies at both low \citep{oes10a} 
and high redshifts \citep{oes10b,gra11,ate15b,fin15} have also assumed a log-normal size distribution. Our size distribution 
assumption in this work is different from A14, where we assumed a normal (not a log-normal) distribution centered at 0.7 kpc with a standard deviation of 0.2 kpc \citep{law12}. \\

 \begin{figure}
\epsscale{2.0}
\includegraphics[trim=0cm 0cm 0cm 0cm,clip=true,width=\columnwidth]{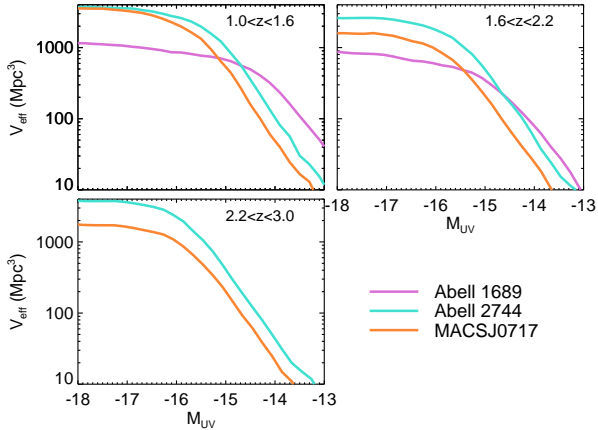}
\caption{The effective volume estimates at each redshift slice in each field. The HFF clusters provide a large volume over
 faint magnitudes ($M_{UV}<-15.5$), while the A1689 cluster enables to probe even fainter galaxies ($M_{UV}>-14.5$) beyond the HFFs magnification limits.}
\label{fig:volumes}
\end{figure}
 
 \begin{figure*}[!t]
\centering
\includegraphics[angle=0,trim=0cm 0cm 0cm 2.5cm,clip=true,width=2.0\columnwidth]{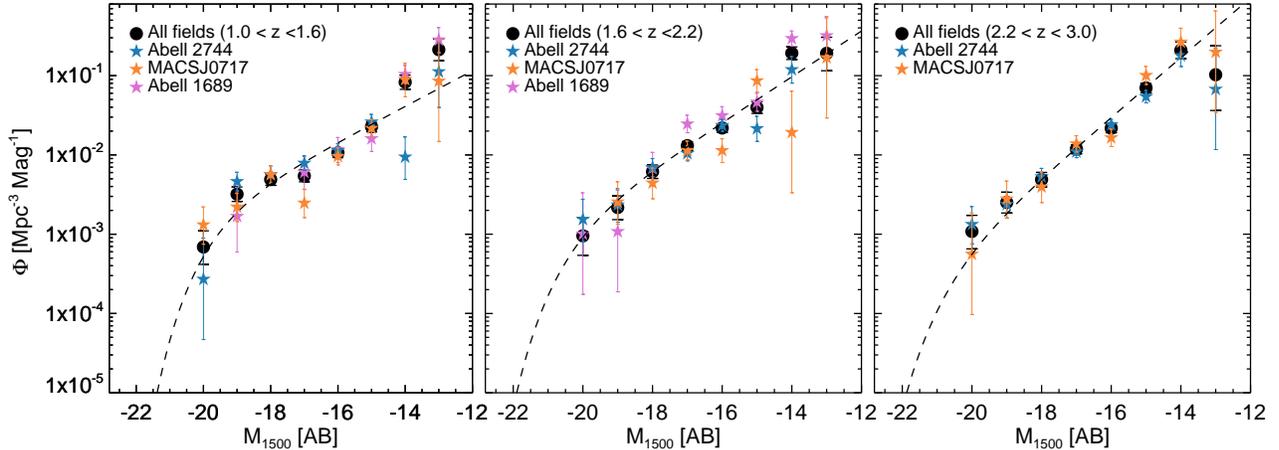}
\centering
\vskip-0.4in
\caption{The rest-frame UV luminosity function for each lensing cluster at $1.0<z<1.6$, $1.6<z<2.2$ and $2.2<z<3.0$ in the left, 
middle and right panels, respectively. The purple, blue and orange stars show the binned LF of A1689, A2744 and MACSJ0717, 
respectively. The black circles are the binned LFs after combining all of the three lensing clusters. The dashed line is the best-fit 
Schechter function (see Section \ref{subsec:mle}).} 
\label{fig:lf_all_redshift}
\end{figure*}

{\it Updating the The Effect of Lensing Magnification and Shear}: The next step in the simulation is to add the lensing effect 
by amplifying the flux and enlarging the size of the galaxies. The way that gravitational lensing distorts the image of a galaxy 
is a combination of convergence (i.e., $\kappa$, stretching a source isotropically) and shear (i.e., $\gamma$, stretching a 
source along a privileged direction). As discussed in other works \citep[e.g.,][]{oes15}, it is crucial to account for the effect 
of lensing distortion. In A14, we did include the effect of convergence in distorting our simulated galaxies. 
For this work, we do a more complete and complex analysis such that the shape of the final distorted image can be 
described using tangential ($\mu_{t}=(1-\kappa-\gamma)^{-1}$) and radial ($\mu_{r}=(1-\kappa+\gamma)^{-1}$) magnification. 
As formulated in \citet{bar10},  a circular source with a circularized radius of $r_{e}$ becomes an elliptical image with semi-major(a) and -minor(b) axises as below: 
  \begin{align}
a=\mu_{r}r_{e}\\
  b=\mu_{t}r_{e} 
   \label{eq:magnification}
 \end{align}
We use the lensing models to construct the $\mu_{t}$ and $\mu_{r}$ maps at source plane for desired redshifts. 
We then use these maps to select random ($\mu_{t} , \mu_{r}$) pairs and distort the image of our simulated galaxies. 
We also increase the flux with a magnification factor of $\mu=\mu_{t}.\mu_{r}$. To mimic the same condition as real galaxies, 
we should note that we exclude the large cluster members from our source plane area reconstruction as the real galaxies 
behind these low-$z$ intervening galaxies can not be observed. 

The corresponding synthetic SED assigned to each simulated source is multiplied by the same filter curves as used in the observations 
to generate artificial catalogs. We then add random photometric noise to the distorted image of each galaxy for each band. 
To detect the galaxies and generate the artificial catalogs, we use the same detection parameters as we used in  \texttt{SExtractor} for our real galaxies.

 Finally, for each cell of the 3-D grid, we have a \texttt{SExtractor} output catalog for 300 artificially created galaxies in random lines 
 of sight, with random sizes and dust attenuation values sampled from the corresponding distributions explained above. 
 We then run the \texttt{EAZY} code on these simulated catalogs and adopt the same selection criteria as we did for the real 
 sources (see Section \ref{sec:selection}). Consequently, we calculate the completeness correction factor, $C(m,z,\mu_{\mathrm{mag}})$, 
 as a function of intrinsic apparent magnitude (i.e., before magnification, m), redshift (z) and magnification ($\mu_{\mathrm{mag}}$) by 
 counting the fraction of recovered artificial galaxies. Figure \ref{fig:Completeness} shows the completeness contours as a 
 function of intrinsic apparent magnitude, m, on the y-axis and redshift on the x-axis for each redshift interval and for each lensing 
 cluster with different colors. The contours are plotted for a magnification of $\mu_{\mathrm{mag}}=2.0$ mag. We can see the 
 difference between HFFs and A1689 completeness values at the lower redshift range ($1.0<z<1.6$), where F225W photometry 
 in A1689 helps to better constrain the redshift and avoids contamination from galaxies with input redshifts below 1.0. As seen 
 in this figure, the recovered redshift distribution from completeness simulations is in agreement with our targeted redshift ranges for each sample.

\subsection{The Effective Survey Volume}
\label{subsec:volume}
We incorporate the completeness corrections in the computation of the effective survey volume, $V_{eff}$, in each magnitude bin as below:
\begin{equation}
V_{\mathrm{eff}}(m)=\int_{0}^{\infty}  \int_{0}^{\infty} \frac{dV_{com}}{dz d\Omega}C(z,m,\mu)\Omega(\mu,z) \mathrm{d}z \mathrm{d}\mu
\label{eq:v_eff}
\end{equation}
where $dV_{com}$ is the comoving volume element at redshift $z$ per unit area, $d\Omega$. In this equation, $C(z,m,\mu)$ 
is the completeness function that depends on redshift (z), intrinsic apparent magnitude (m) and magnification ($\mu$). $\Omega(\mu,z)$ is 
the area element in the source plane at z which is magnified by a factor of $\mu$. We run \texttt{Lenstool} for each 
aforementioned cluster mass model to generate the de-lensed magnification maps at different redshifts. 
We then use these maps to estimate the $\Omega(\mu,z)$ of each cluster at each redshift. Similar to our completeness 
simulations (see Section \ref{sec:completeness}), we subtract the area occupied by the large cluster members from our source plane area reconstruction.

Figure \ref{fig:volumes} represents the effective volumes versus the absolute magnitude at 1500 \AA, $M_{UV}$, for 
each cluster at three redshift ranges. This plot clearly shows the importance of including A1689 for finding the faintest galaxies ($M_{UV}>-14.5$). 
We should emphasize that the small volumes at faint luminosities are not necessarily due to a large incompleteness but because of small 
area available at these magnitudes. For the volume calculation at each magnitude, unlike the field studies where the full area is available, here only 
a portion of area (i.e., effective area) with enough magnification (i.e., minimum magnification required for detection at each magnitude) is used. 
Therefore, at very faint luminosities, only a tiny fraction of area is available for the volume measurements.

\section{Luminosity Function of Photometric Redshift Samples}
\label{sec:lf_photz}
Using the effective volumes, we construct the UV luminosity function of our photometric redshift selected galaxies at the 
peak epoch of cosmic star formation rate density. To be consistent with other studies at the same redshift ranges \citep[e.g.,][]{oes10a, par16} 
and at higher redshifts \citep[e.g., ][]{bou07,fin15}, we measure the UV luminosities at rest-frame 1500 \AA. 

The galaxy luminosity function is commonly fitted by a Schechter function \citep{sch76} characterized by an exponential 
behavior at luminosities brighter than a characteristic magnitude, $M^{*}$, and a power-law at the faint end with slope $\alpha$ as below:
\begin{equation}
\phi(M)=0.4\mathrm{ln}(10)\phi^{*}10^{-0.4(M-M^{*})(1+\alpha)}e^{-10^{-0.4(M-M^{*})}}
\end{equation}
where $\phi^{*}$ is the normalization of this function.

In this section, we first calculate and compare binned UV LFs of each cluster field and then we find the best-fit Schechter
 parameters for the combined LF using a maximum likelihood approach on the unbinned data.

\subsection{The Binned UV LFs}
\label{subsec:binned lf}
The LF at each $M_{1500}$ bin is derived using the measured $V_{eff}$ values which account for the completeness corrections. 
This is the commonly used $V_{eff}$ method \citep[e.g., A14, ][]{oes10a} where one calculates the number density of galaxies in 
each bin by dividing the number of galaxies in the corresponding absolute magnitude bin by the effective volume of that bin. 
But the effective volume might change significantly from one side of the magnitude bin to the other. Therefore, we estimate the
 effective volume for each individual galaxy and then sum up over all the galaxies within each bin, as shown below:
\begin{equation}
\phi(M_{i})dM_{i}=\sum_{j=1}^{N}\frac{1}{V_{\mathrm{eff}}(M_{j})}
\end{equation}

As illustrated in Figure \ref{fig:lf_all_redshift}, we estimate the binned LFs of each lensing field separately as well as a total LF 
combining all of the cluster fields. For the combined LF,  the $V_{eff}$ is a sum of the effective volumes over all of the cluster fields. 

For each bin with a large number of galaxies ($N>50$), we assign an uncertainty of $\frac{\phi_{i}}{\sqrt{N}}$ using Poisson statistics. 
In the case where less than 50 galaxies are in the bin, we compute the Poisson approximation, $\Delta_{P}$, from \citet{geh86} and 
assign an uncertainty of $\frac{\phi_{i} \Delta_{P} }{N}$ to each bin. Each bin has a width of $\Delta M_{UV}=1$ magnitude and our 
faintest magnitude bin is centered at $M_{UV}=-13$ (i.e., a magnitude cut at $M_{UV}=-12.5, $ see Section \ref{sec:selection}). 
The values of the binned LFs, and the number of galaxies at each bin are listed in Table \ref{tab:LF_bin}.

The binned LFs are good for visualization but poor for inference because of arbitrary bin widths, bin centers and loss of information
 within each bin. Therefore, instead of using binned estimators, we use an unbiased, unbinned maximum likelihood estimator as explained in the next section.

\begin{deluxetable}{cccc}
\tablecaption{Binned UV LFs}
\tabletypesize{\footnotesize}
\tablewidth{1\columnwidth}
\tablehead{\colhead{z} & \colhead{$M_{UV}$} & \colhead{Number of sources} & \colhead{$\phi$ ($\times 10^{-2}$Mpc$^{-3}$mag$^{-1}$)}\\
\cline{1-4}\\
\multicolumn{4}{c}{Photometric redshift  LFs}}
\startdata $1.0<z<1.6$    & -20.0  &      6           &      0.069$\substack{+0.041 \\ -0.027}$\\
				       &  -19.0  &     27           &      0.320$\substack{+0.074 \\ -0.061}$\\
				       &  -18.0  &     41           &      0.490$\substack{+0.089 \\ -0.076}$\\
				       &  -17.0  &     41           &      0.543$\substack{+0.099 \\ -0.084}$\\
				       &  -16.0  &     60           &      1.064$\substack{+0.137 \\ -0.137}$\\
				       &  -15.0  &      54           &      2.266$\substack{+0.308 \\ -0.308}$\\
				       &  -14.0  &      28           &      8.275$\substack{+1.877 \\ -1.554}$\\
				       &  -13.0  &      13           &      21.279$\substack{+7.693 \\ -5.826}$\\  
\hline
                 $1.6<z<2.2$    & -20.0   &      5          &      0.095$\substack{+0.064 \\ -0.041}$\\
				       & -19.0  &     11         &      0.217$\substack{+0.087 \\ -0.064}$\\
				       & -18.0  &     31         &      0.615$\substack{+0.131 \\ -0.110}$\\
				       & -17.0  &     62         &      1.307$\substack{+0.166 \\ -0.166}$\\
				       & -16.0  &     69         &      2.186$\substack{+0.263 \\ -0.263}$\\
				       & -15.0  &     40         &     3.964$\substack{+0.731 \\ -0.624}$\\
				       & -14.0  &     35         &      19.223$\substack{+3.828 \\ -3.235}$\\
				       & -13.0  &      6          &      19.104$\substack{+11.412 \\ -7.578}$\\  					
\hline
                $2.2<z<3.0$        & -20.0 &      6          &      0.108$\substack{+0.065 \\ -0.043}$\\
				         & -19.0 &    14          &      0.252$\substack{+0.087 \\ -0.066}$\\
					&  -18.0 &   27          &      0.488$\substack{+0.113 \\ -0.093}$\\
					&  -17.0 &   61          &      1.186$\substack{+0.152 \\ -0.152}$\\
					&  -16.0 &   67          &      2.170$\substack{+0.265 \\ -0.265}$\\
					&   -15.0 &   53         &      7.003$\substack{+0.962 \\ -0.962}$\\
					&  -14.0 &    21         &      20.915$\substack{+5.637 \\ -4.532}$\\
					&  -13.0 &    2           &      10.271$\substack{+13.547 \\ -6.635}$\\  				
\cutinhead{LBG LFs} 
		$z\sim1.65$ 	 &  -18.0  &     3           &      0.527$\substack{+0.513 \\ -0.287}$\\
				   	 &  -17.0  &   4             &      0.733$\substack{+0.579 \\ -0.351}$\\
				   	 &  -16.0  &    5            &      1.329$\substack{+0.899 \\ -0.574}$\\
				   	 &  -15.0  &     2           &      1.150$\substack{+1.517 \\ -0.743}$\\
				 	 &  -14.0  &    3           &      8.566$\substack{+8.332 \\ -4.663}$\\
				  	 &  -13.0  &     1           &      10.477$\substack{+24.098 \\ -8.665}$\\   
\hline
		$z\sim2.0$	 & -20.0  &        3            &      0.058$\substack{+0.057 \\ -0.032}$\\
				  	 & -19.0 &        10           &      0.199$\substack{+0.085 \\ -0.062}$\\
				     	 & -18.0    &      34           &      0.768$\substack{+0.156 \\ -0.131}$\\
				     	 &  -17.0    &     50           &      1.617$\substack{+0.263 \\ -0.228}$\\
				     	 &  -16.0   &      40           &      3.556$\substack{+0.656 \\ -0.560}$\\
				    	 &   -15.0   &      20           &     5.736$\substack{+1.592 \\ -1.271}$\\
				          &  -14.0  &        13           &      20.106$\substack{+7.269 \\ -5.504}$\\
					 & -13.0   &          2           &     25.922$\substack{+34.192 \\ -16.746}$\\  		
\hline
		$z\sim2.7$ 	& -20.0   &      6          &      0.112$\substack{+0.067 \\ -0.045}$\\
				  	& -19.0   &      10        &      0.186$\substack{+0.079 \\ -0.058}$\\
				     	& -18.0    &      21       &      0.409$\substack{+0.110 \\ -0.089}$\\
				     	&  -17.0   &      46       &      1.309$\substack{+0.223 \\ -0.192}$\\
				     	&  -16.0   &      37       &      4.218$\substack{+0.814 \\ -0.691}$\\
				    	&   -15.0  &      14       &     7.796$\substack{+2.690 \\ -2.060}$\\
				         &  -14.0    &      5         &      32.106$\substack{+21.716\\ -13.870}$\\
					& -13.0    &       1         &    24.187$\substack{+55.631 \\ -20.003}$ 
\enddata
\label{tab:LF_bin}
\end{deluxetable}

\begin{figure*}[!t]
\centering
\includegraphics[angle=0,trim=0cm 0cm 0cm 2.5cm,clip=true,width=2.\columnwidth]{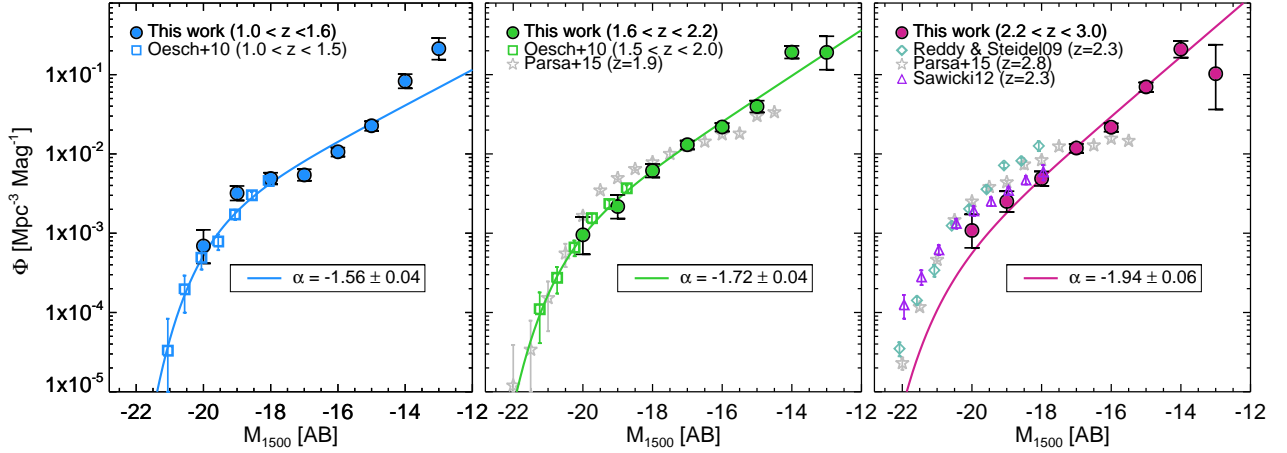}
\centering
\vskip-0.2in
\caption{Rest-frame UV luminosity functions at $z\sim1.3$ (left), $z\sim1.9$(middle) and $z\sim2.6$(right). The blue, green and red 
circles are our binned LFs combining all three lensing clusters (see Section \ref{subsec:binned lf}). The blue and green squares are 
the LFs from \citet{oes10a} at $1.0<z<1.5$ and $1.5<z<2.0$, respectively, of which the individual data were used for our MLE fitting. 
The light blue diamonds are the LFs from \citet{red09} based on a BX selected sample of star-forming galaxies. The gray stars and 
purple triangles are the results from \citet{par16} (photometric redshift selection) and \citet{saw12} (BX selection), respectively. 
The solid line in all three panels shows our best Schechter fit through a MLE technique.} 
\label{fig:lf}
\end{figure*}

\begin{deluxetable*}{ccccc}
\tablecaption{Best-fit Schechter Parameters for UV LFs}
\tabletypesize{\footnotesize}
\tablewidth{2\columnwidth}
\tablehead{\colhead{z} & \colhead{$\alpha$} & \colhead{M$^{*}$} & \colhead{$\phi^{*}$($10^{-3}$Mpc$^{-3}$ mag$^{-1}$)}\\
\cline{1-5} \\
\multicolumn{4}{c}{Photometric redshift LF, MLE fitting}}
\startdata $1.0<z<1.6$\tablenotemark{a}      &        -1.56$\pm$0.04         &     -19.74$\pm$0.18        &          2.32$\pm$0.49\\
                $1.6<z<2.2$\tablenotemark{a}      &        -1.72$\pm$0.04 	   &     -20.41$\pm$0.20        &          1.50$\pm$0.37\\
                $2.2<z<3.0$\tablenotemark{b}      &        -1.94$\pm$0.06          &     -20.71$\pm$0.11(prior) &       0.55$\pm$0.14\\
\cutinhead{LBG LF, $\chi^{2}$ fitting} 
                      $z\sim1.65$\tablenotemark{c}             &          -1.50$\pm$0.16        &     -19.85$\pm$0.41       &           2.21$\pm$1.32\\
                      $z\sim2.0$\tablenotemark{d}               &          -1.80$\pm$0.06        &     -20.39$\pm$0.31       &            1.46$\pm$0.65\\
                      $z\sim2.7$\tablenotemark{e}                &          -2.01$\pm$0.08        &      -20.70(fixed)             &              0.48$\pm$0.15
\enddata
\tablenotetext{a}{Maximum likelihood fit to the whole sample including individual galaxies from all three lensing clusters as well as the bright-end galaxies from \citet{oes10a}.}
\tablenotetext{b}{Maximum likelihood fit to the individual galaxies from the HFF clusters assuming a Gaussian prior for $M^{*}$ (see Section \ref{subsec:mle})}
\tablenotetext{c}{$\chi^{2}$ fitting to the binned data from A1689 as well as the bright-end LBGs from \citet{oes10a}}
\tablenotetext{d}{$\chi^{2}$ fitting to the binned data from all three lensing clusters as well as the bright-end LBGs from \citet{oes10a}}
\tablenotetext{e}{$\chi^{2}$ fitting to the binned data from the HFF clusters assuming a fixed $M^{*}$ (see Section \ref{sec:lf-lbg}).}
\label{tab:LF_param}
\end{deluxetable*}
  
\begin{figure}[!t]
\centering
\includegraphics[angle=0,trim=0.2cm 0cm 0.5cm 0cm,clip=true,width=0.85\columnwidth]{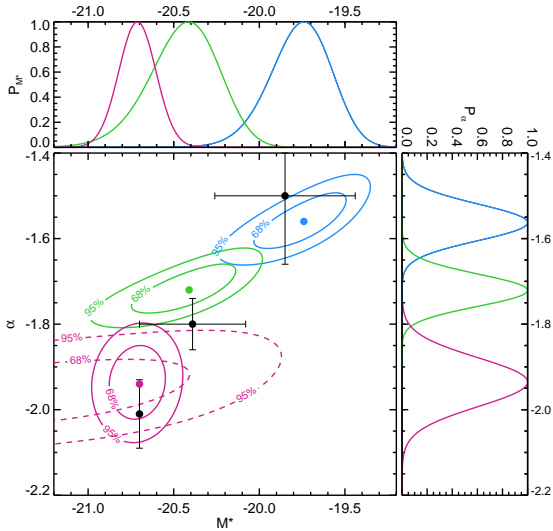}
\centering
\caption{ The $68\%$ and $95\%$ contours of the $z\sim1.3$, $z\sim1.9$ and $z\sim2.6$ photometric redshift LFs are 
shown with blue, green and red colors, respectively. The red dashed line shows the contours for $z\sim2.6$ LF before adding 
the $M^{*}$ prior (see Section \ref{subsec:mle}). The marginalized probability distribution of each parameter $P(\alpha)$ 
and $P(M^{*})$ are also plotted on the right and top sides, respectively. The best-fit values of Schechter parameters for 
each LF is shown with a filled circle. The black filled circles with error bars denote the best-fit values for the LBG LFs (see Section \ref{sec:lf-lbg}).} 
\label{fig:lf_contour}
\end{figure}

\subsection{The Unbinned Maximum Likelihood Estimator}
\label{subsec:mle}
In this section, we explain our methodology to estimate the best Schechter function parameters by maximizing the likelihood 
function of the unbinned data. The standard maximum likelihood (MLE) technique was first used by \citet[STY79]{san79}, 
and later by many other studies to derive the best-fit parameters for UV LFs at intermediate redshifts (A14), high
 redshifts \citep[e.g., ][]{mcl13,bou15} and for the H${\alpha}$ LF \citep{meh15}. Here, we adopt a similar approach as in 
 A14 where we modify the standard STY79 MLE technique to account for uncertainties in the measurements of the absolute 
 magnitude. This modified methodology is also used in \citet{meh15}. In the MLE technique, the best fit is found through maximizing 
 the joint likelihood function $L$ defined as below.
 
 \begin{equation}
 \mathcal{L}=\prod_{i=1}^{N}P(M_{i})
  \label{eq:L}
 \end{equation}
where in the standard MLE, $P(M_{i})$ defined as below:
\begin{equation}
P(M_{i})=\frac{\phi(M_{i}) V_{\mathrm{eff}}(M_{i})}{\int_{-\infty}^{M_{\mathrm{limit}}} \phi(M) V_{\mathrm{eff}}(M) \mathrm{d} M}
 \label{eq:pm}
\end{equation}

where N is the total number of objects in each sample. $P(M_{i})$ is the probability of finding a galaxy with absolute 
magnitude $M_{i}$ in a corresponding effective volume, $V_{eff}(M_{i})$. We calculate this probability value for all of the galaxies 
in each of our samples. $\phi(M_{i})$ is the parametric luminosity function assuming a Schechter function. The $M_{limit}$ is 
defined for each sample to be the faintest absolute magnitude (i.e., corrected for the magnification). The $M_{limit}$ values 
are -12.88, -12.12 and -13.40 for z=1.3, 1.9 and 2.6 samples, respectively. 

To incorporate absolute magnitude uncertainties in the LF analysis, we assume a Gaussian probability distribution 
$G(M|M_{i},\sigma_{i})$ for each object centered at the object's absolute magnitude $M_{i}$ and a standard deviation equal to
 the object's absolute magnitude uncertainty $\sigma_{i}$. We then modify the Equation \ref{eq:pm} as below:

\begin{equation}
P(M_{i})=\frac{\int_{-\infty}^{+\infty} \phi(M) V_{\mathrm{eff}}(M) G(M|M_{i},\sigma_{i}) \mathrm{d} M}{\int_{-\infty}^{M_{\mathrm{limit}}} \phi(M) V_{\mathrm{eff}}(M) \mathrm{d} M}
 \label{eq:pm_mod}
\end{equation}

with
\begin{equation}
G(M|M_{i},\sigma_{i})=\frac{1}{\sqrt{2\pi}\sigma_{i}}exp(-\frac{(M-M_{i})^{2}}{2\sigma_{i}^{2}})
\end{equation}

As also considered in A14, for our lensed galaxies, the total uncertainty, $\sigma_{i}$, of intrinsic absolute magnitude is due 
to the uncertainty in photometric measurements ($\sigma_{m}$), photometric redshifts ($\sigma_{z}$) and the lens models 
($\sigma_{model}$).  Below, we investigate in detail these different sources of uncertainties.
\begin{enumerate}[label=\alph*.]

\item $\sigma_{m}$: The photometric uncertainties are calculated using the \texttt{SExtractor} output of flux uncertainties.

\item $\sigma_{z(total)}$: The photometric redshift uncertainty, $\sigma_{z}$, for each galaxy is computed as 1$\sigma$ confidence 
interval of its redshift probability distribution from \texttt{EAZY}. This redshift uncertainty impacts the measured intrinsic absolute 
magnitude in two ways. First, since the distance modulus is dependent on the redshift, we estimate the effect of redshift uncertainty 
on the absolute magnitude through an error propagation of Equation \ref{eq:M}. Second, the magnification value of each galaxy 
is estimated through running \texttt{Lenstool} while incorporating its photometric redshift as an input. Therefore, a redshift uncertainty 
causes a magnification uncertainty, $\sigma_{\mu(z)}$. To estimate $\sigma_{\mu(z)}$ for each galaxy, we run \texttt{Lenstool} 
for 100 random redshifts generated from a Gaussian redshift distribution centered at the galaxy's photometric redshift with standard 
deviation equal to $\sigma_{z}$. The distribution of output random magnifications for each galaxy is fitted with a Gaussian function 
to derive $\sigma_{\mu(z)}$. Because $\sigma_{z}$ and $\sigma_{\mu(z)}$ are correlated, we calculate the total redshift uncertainty 
as a sum over them, $\sigma_{z(total)}=\sigma_{z}+\sigma_{\mu(z)}$
\item $\sigma_{\mu(model)}$: The final source of uncertainty is related to the lensing models. To estimate this uncertainty, 
we randomly sample the parameter space of each lens model. A detailed description of these measurements is given in A14.
\end{enumerate}
 We calculate the total uncertainty of intrinsic absolute magnitude by adding all these uncertainties in quadrature. 

Substituting Equation \ref{eq:pm_mod} in Equation \ref{eq:L}, we calculate the likelihood function over a grid of faint-end 
slope ($\alpha$) and characteristic magnitude ($M^{*}$). The small survey areas probed in this study, limits the number 
of bright galaxies (i.e., $M<M^{*}$). Therefore, to constrain $M^{*}$, we combine our $z\sim1.3$ and $z\sim1.9$ samples 
with the samples from a wider survey from \citet{oes10a}. To be consistent with our samples, we use their photometric 
redshift selected galaxies at $1.0<z<1.5$ and $1.5<z<2.0$. 

For our $2.2<z<3.0$ sample, our brightest LF bins are lower than the values from the literature. 
Furthermore, we do not have access to the individual galaxies from the literature.
Therefore, we adopt a different approach to find the best-fit LF. We multiply the likelihood function by an $M^{*}$ prior to 
compute the posterior function. Utilizing the best Schechter parameters reported in \citet{red09}, we define the prior as a 
Gaussian function centered at -20.70 with standard deviation of 0.11. 
We should note that this discrepancy between the LFs at bright luminosities, is not due to our completeness correction, as our $z\sim2.6$ sample is 
$> 90\%$ complete at these luminosities (see Figures \ref{fig:Completeness} and \ref{fig:volumes}). 
Considering that we only have two clusters at this redshift range, and consequently we probe a small area, 
it is not unlikely that this low number density may be due to a presence of an underdense region of galaxies. 
The reason that this under-density appears to be affecting the bright end more than the faint end can be 
understood by different spatial clustering of the bright galaxies relative to the faint ones \citep[e.g., ][]{zeh05}.\footnote{In the future, 
when we complete the UV survey of HFFs, we will add 4 more clusters and consequently triple our sample size at $z=2.6$. 
Therefore, our number density measurement for bright galaxies at this redshift will be more accurate.}

\begin{figure*}[!t]
\centering
\includegraphics[angle=0,trim=0cm 0cm 0cm 2.5cm,clip=true,width=2.0\columnwidth]{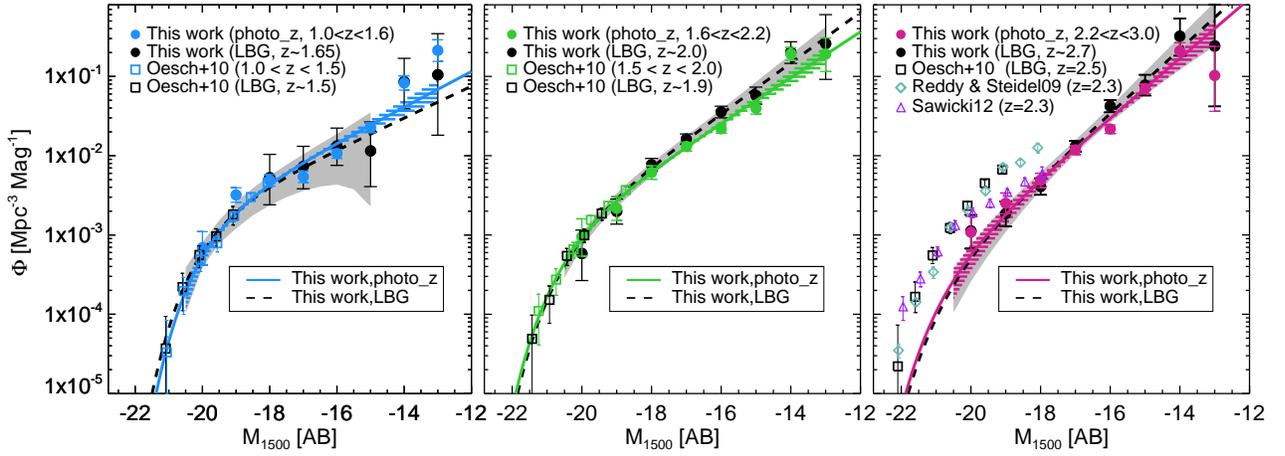}
\centering
\vskip-0.4in
\caption{Comparing the UV LF of the LBG and photometric redshift samples. The black filled circles are the binned LBG LF for 
our F225W-, F275W- and F336W-dropout samples from left to right panels, respectively. The black dashed line in each panel 
represents the best-fit Schechter function for the corresponding LBG LF. The gray regions indicate the $\pm2\sigma$ confidence 
region for each fit derived via Monte Carlo simulation. We over-plot the binned and best-fit LFs derived from our photometric
 redshift samples as shown before in Figure \ref{fig:lf}. The hatched region denotes the $\pm2\sigma$ confidence region of the 
 best-fit photometric redshift LF. The black open squares are the binned LBG LFs from \citet{oes10a}. The rest of the colors and 
 symbols are as in Figure \ref{fig:lf}. For each redshift range, our binned and best-fit LFs are in consistent within the $2\sigma$ error bars.} 
\label{fig:lf-lbg}
\end{figure*}

The best estimates of Schechter parameters are derived via marginalization of posterior functions at all redshifts. 
Figure \ref{fig:lf} shows the binned LFs along with our best MLE  determinations at each redshift range. The best Schechter 
parameters are tabulated in Table \ref{tab:LF_param}. Our MLE estimates reveal steep faint-end slopes 
of $\alpha=-1.56\pm0.04$, $-1.72\pm0.04$ and $-1.94\pm0.06$ for $z\sim1.3$, $z\sim1.9$ and $z\sim2.6$ samples, respectively. 
We emphasize that our estimate of the faint-end slope at $z\sim2.6$ is mostly independent of our choice of the $M^{*}$ prior, 
as we derive $\alpha=-1.97\pm0.06$ in the absence of a prior. Our steep LFs show no sign of turnover down to $M_{UV}=-12.5$ mag.

The contours in Figure \ref{fig:lf_contour} illustrate the correlation between the faint-end slope ($\alpha$) and the 
characteristic magnitude ($M^{*}$). The best Shechter parameters derived through marginalization are shown with filled blue, 
green and red circles for $z\sim1.3$, $z\sim1.9$ and $z\sim2.6$ samples, respectively. We are also overplotting our 
best LFs (filled black circles) from LBG samples (see Section \ref{sec:lf-lbg}). The red dashed contours show the Likelihood function at $z\sim2.6$, before incorporating the $M^{*}$ prior. 

The systematic uncertainties -- particularly in the size distribution assumption at faint luminosities -- may affect 
 the completeness corrections and thus the LF measurements at these magnitudes. This concern is also expressed in a recent paper by \citet{bou16}, where they 
 measure very small sizes (160-240 pc) for ultra-faint galaxies ($M_{UV}=-15$) at $2<z<8$ and then discuss the possible effects 
 due to uncertain size assumptions on the LF measurements. We should emphasize that they present their conclusions for a large redshift range of $z=2-6$, 
 while we expect the lower redshift galaxies ($z\sim2$) to be on average larger than their high redshift counterparts (as seen at higher luminosities, \citet{shi15}). 
 Our assumed size distribution for ultra-faint galaxies is the closest to the \citet{bou16} measurements, relative to the other LF studies. We run some experiments to investigate whether 
 the faint galaxies with large completeness corrections (i.e., where the systematic uncertainty dominates), are dictating our best-fit LFs by excluding 
 all of our galaxies with completeness below $50\%$. This reduces the size of our $z\sim1.3$, $z\sim1.9$ and $z\sim2.6$ samples by 33\%, 53\% and 44\%, 
 such that our final ``complete" samples have 186, 127 and 141 galaxies, respectively. To be consistent, we also remove the corresponding volumes 
 from our total volume estimates. We then re-fit the LFs and measure faint-end slopes of $\alpha=-1.55\pm0.06$, $\alpha=-1.69\pm0.07$ and $\alpha=-1.79\pm0.08$ 
 at $1<z<1.6$, $1.6<z<2.2$ and $2.2<z<3.0$, respectively. 
 These estimates are all steep and show the same trend of steeper slopes toward higher redshifts, though with slightly shallower slopes. We note that, although the $z\sim1.3$ and $z\sim1.9$ faint-end slopes measured from the ``complete" sample are consistent with the slopes measured from the full sample, the $z\sim2.6$ slope from the ``complete" sample is significantly shallower by $\sim1.5\times$ the individual errors added in quadrature (though the measurements aren't completely independent, so adding in quadrature will slightly overestimate the uncertainties). The probability of obtaining such a deviation in at least one of the three slope measurements is small (10\%), 
 and suggests that the systematic uncertainties are not negligible. 
 Consequently, as also emphasized in \citet{bou16}, the size measurements of very faint galaxies 
  will need to be more accurately determined for higher quality LF measurements.

\section{Luminosity Function of LBG Samples}
\label{sec:lf-lbg}
As discussed in Section \ref{introduction}, one of the goals of the present paper is to understand the effect of two widely 
used selection techniques. To this end, we have also performed a parallel determination of the UV luminosity function based 
on the Lyman break ``dropout" selection at equivalent redshift ranges. A complete description of our color-color selection, 
sample contamination and the completeness simulation for dropouts is given in Appendix \ref{appendixA}. As explained there, 
our LBG samples consist of 19 F225W-, 178 F275W- and 142 F336W-dropouts at $z\sim1.65$, $z\sim2.0$ and $z\sim2.7$, 
respectively. We note that our LBG samples have fewer galaxies than our photometric redshift samples, because we 
require $5\sigma$ detection in the detection filter for these samples (see Appendix \ref{appendixA}), whereas the photometric 
redshift samples only require a $3\sigma$ detection (see Section \ref{sec:selection}). 
To ensure accurate detection of a break, we restrict our sample to objects where the imaging depth is sufficient to detect 
at least a one magnitude break (at 1 sigma) between the dropout filter (F225W, F275W, and F336W at $z\sim$1.65, 2.0, and 2.7, respectively) 
compared to the adjacent longer wavelength filter. This cut only removes two galaxies from the A2744 F336W-dropout LF and 
it does not change the rest of the LBG samples. 

The effective volume including the completeness corrections is calculated for these samples using Equation \ref{eq:v_eff}. 
In order to estimate the binned UV LF for our LBG samples, we use the same methodology as we used for our primary photometric 
redshift samples. Similarly, we restrict our dropout samples to galaxies with $M_{1500}<-12.5$ for the same reasons that were 
mentioned before (see Section \ref{sec:selection}). This limit excludes 1 ($\sim5.3\%$), 6 ($\sim3.4\%$) and 0 galaxies 
from the F225W-, F275W-, F336W-dropout LFs, respectively. Finally, we have 18, 172 and 140 galaxies for the $z\sim1.65$, $z\sim2.0$ and $z\sim2.7$ LBG LFs, respectively.

Furthermore, to constrain the bright-end of our F225W and F275W-dropout LFs, we incorporate the binned measurements 
from \citet{oes10a} LBG samples. Here, we do not use the MLE technique because we do not have individual measurements for 
all of these bright-end LBG samples. We determine the best Schechter parameters only using the simple $\chi^{2}$ technique, 
considering that these two methods of fitting (MLE vs $\chi^{2}$) show good agreement for the photometric redshift LFs.
For our F336W-dropout LF, we only fit to our binned data keeping the characteristic magnitude $M*$ at a fixed value of -20.7, 
similar to what we used for our $z\sim2.6$ photometric redshift LF. The binned values and the best-fitting Schechter parameters 
for the LBG LFs are given in Tables \ref{tab:LF_bin} and \ref{tab:LF_param}, respectively.

\section{Discussion}
\label{sec:discussion}
\subsection{Comparing the UV LFs of Photometric redshift and LBG Samples }
\label{sec:photz_vs_lbg}
Figure \ref{fig:lf-lbg} compares our LF results derived for the photometric redshift and UV-dropout selections. From left to right, 
the F225W-, F275W- and F336W-dropout LFs shown with black circles are compared with the photometric redshift 
LFs at $z\sim1.3$ (blue circles), $z\sim1.9$ (green circles) and $z\sim2.6$ (red circles), respectively. Together with our 
data points for each redshift range, we also show the bright-end LFs of \citet{oes10a} derived from their photometric 
redshift (blue, green and red squares) and UV-dropout (black squares) samples. In addition, we include the LF results from 
several relevant studies \citep{red09,saw12,par16}. To compare these LF measurements, we run a set of Monte 
Carlo simulations and estimate the $2\sigma$ confidence interval from each best-fit Schechter function. The gray and hatched 
regions in Figure \ref{fig:lf-lbg} encompass the $2\sigma$ uncertainties of the LBG and photometric redshift LFs, respectively. 
Because our LBG samples have fewer galaxies than the photometric redshift samples, the corresponding LFs are more uncertain. 
Our LFs are in agreement within these confidence regions. Indeed, similar agreement between LFs derived from these 
two selection techniques at higher redshifts has been shown before \citep{mcl13,sch13}. However, the lack of a robust 
knowledge of various systematic effects such as intrinsic size distribution and dust reddening at these 
faint luminosities still introduces moderate differences between these two LF measurements. 
 
\begin{figure}[!t]
\centering
\includegraphics[angle=0,trim=0cm 2cm 0.5cm 3cm,clip=true,height=12cm,width=1.\columnwidth]{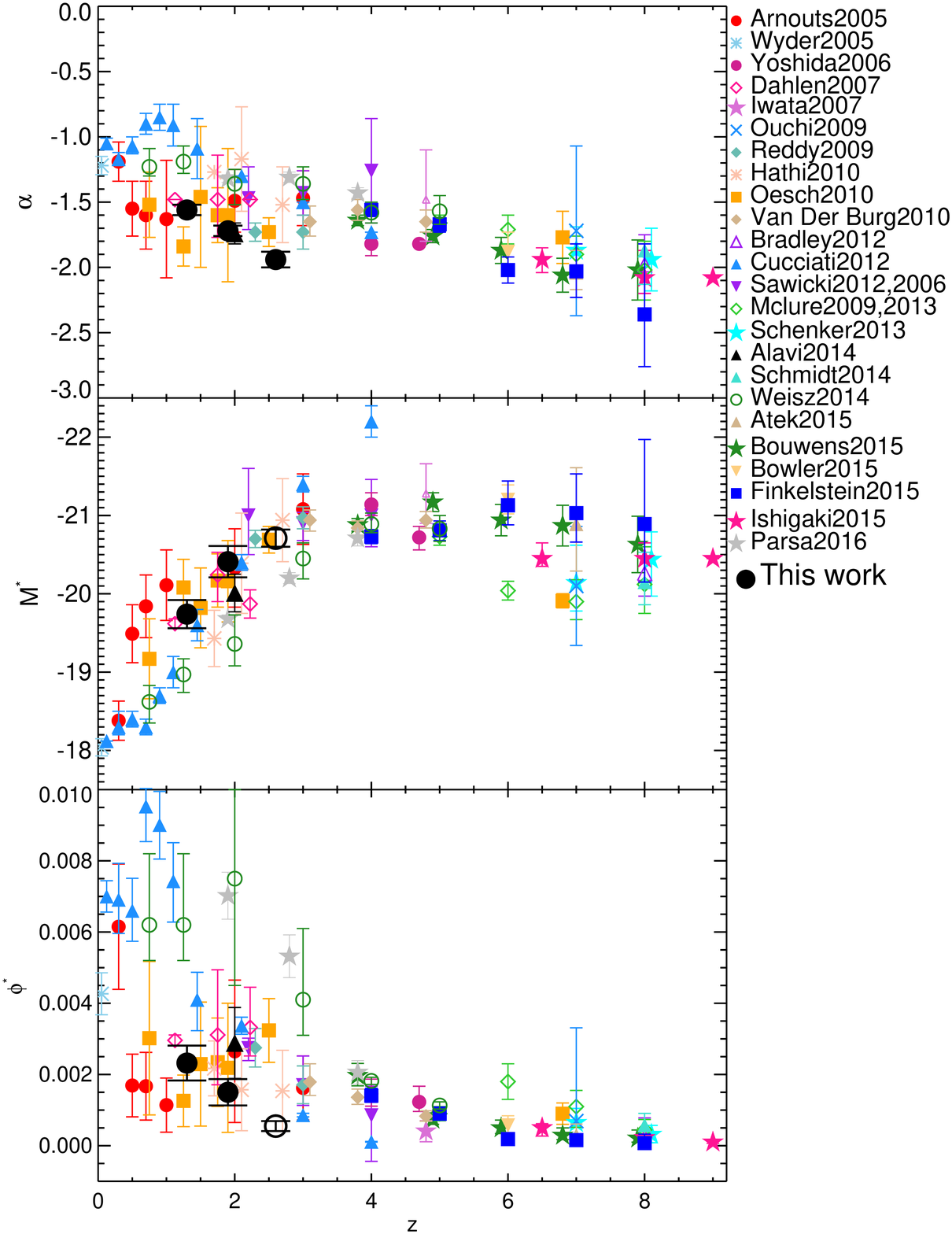}
\centering
\caption{The redshift evolution of the Schechter parameters $\alpha$(top), $M^{*}$(middle) and $\phi^{*}$(bottom). 
The present determinations are shown with black filled circles. Our measurements of $M^{*}$ and $\phi^{*}$ at $z\sim2.6$ 
are shown with black open circles, as they are dependent on our choice of the $M^{*}$ prior. All symbols from the literature 
are summarized in the right-hand side of the plot. A detailed description about each parameter evolution is given in the text (see Section \ref{sec:schechter-evolv}). } 
\label{fig:lf-parameters}
\end{figure}

\begin{figure*}[!t]
\centering
\includegraphics[angle=0,trim=1cm 0cm 1cm 0cm,clip=true,width=2\columnwidth]{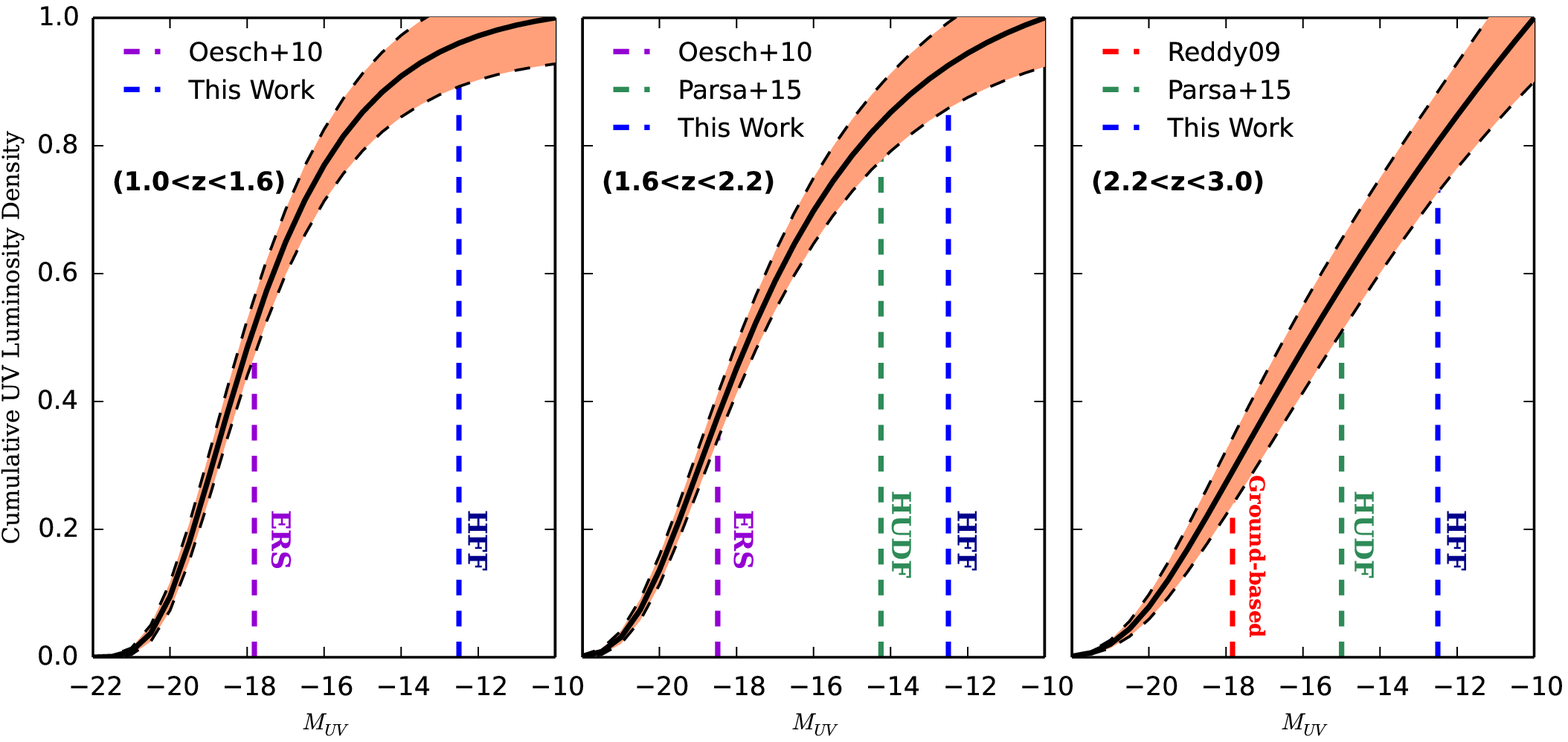}
\centering
\caption{The cumulative UV luminosity density at $1.0<z<1.6$ (left), $1.6<z<2.2$ (middle) and $2.2<z<3.0$ (right). 
The purple, green and blue dashed lines show the UV limiting magnitudes for the ERS \citep{oes10a}, Hubble Ultra Deep Field \citep{par16} 
and our samples. The orange region represents the 1$\sigma$ uncertainty measured at each $M_{UV}$. We have measured the LF 
of galaxies responsible for $58\%$, $55\%$ and $59\%$ of total UV luminosity density at $z\sim1.3$, $z\sim1.9$ and $z\sim2.6$, respectively.} 
\label{fig:lf_all_redshift_nocluster}
\end{figure*}

\subsection{Evolution of the LF Schechter parameters}
\label{sec:schechter-evolv}
In order to understand the evolution of luminosity function parameters with redshift, we compare our best-fit Schechter parameters 
with other determinations of the rest-frame UV luminosity function at higher and lower redshifts in Figure \ref{fig:lf-parameters}. 
We summarize the evolution of LF parameters as below:

Exploiting the magnification from strong gravitational lensing and consequently extending the UV LF to very low luminosities, enables 
a robust estimate of the faint-end slope. In the context of recent UV LF studies, there are not as many measurements 
at $z=1-1.5$ to compare with our estimates, but as can be seen in Figure \ref{fig:lf-parameters}, our inferred value of the 
faint-end slope ($\alpha=-1.56\pm0.04$) at $z\sim1.3$ is consistent with other results from \citet{arn05} and \citet{oes10a} 
given their large uncertainties. We should note that we are in better agreement with the \citet{oes10a} estimate for their LBG 
LF (at $z\sim1.5$), as their photometric redshift LF has a very steep faint-end slope. Regarding our estimate for the $z\sim1.9$ 
LF, we are again in good agreement with several other results, particularly both LBG and photometric redshift LFs of \citet{oes10a} 
and also the $z=2.3$ LF from \citet{red09}. We note that we are also in agreement with our previous $z\sim2$ LBG LF 
from A14 ($\alpha=-1.74\pm0.08$). Finally regarding our estimate for the $z\sim2.6$ LF, we derive a faint-end slope steeper 
than previous determinations and more similar to the steep faint-end slopes favored at higher redshifts. As a consequence, 
we conclude a rapid evolution in the faint-end slope toward shallower values during the 2.2 Gyr from $z=2.6$ to $z=1.3$ 
which seems to continue to $z=0$. We also refer the reader to a recent work by \citet{par16} (see gray filled stars in 
Figure \ref{fig:lf-parameters}) who study the UV LF between $z=2-4$. For both $z=1.9$ and $z=2.8$, they derive a 
value of $\alpha=-1.32\pm0.04$ which is considerably shallower 
than most of the other studies including ours. Consequently, they derive fainter $M^{*}$ and larger $\phi^{*}$ values relative to 
all of the other works at $z=2-3$ in the literature. We note that they do not use the filter that samples the Lyman 
break at $z\sim2$. 

In addition to the observed LFs, we compare our results with the LFs from local group (LG) fossil 
records by \citet{wei14} (open green circles in Figure \ref{fig:lf-parameters}). Using the SFHs of LG galaxies, they reconstruct 
the UV LFs down to very faint magnitudes of $M_{UV}\sim-1.5$. Comparing to our results, they estimate shallower 
faint-end slopes ($\alpha>-1.4$) for their $z=1.25$ and 2.0 LFs, but they derive steeper faint-end slopes when they 
restrict their calculations to the luminosities where their data are complete. 
Although, for an exact comparison, we need to consider the different uncertainties (i.e, small sample size) and systematic 
errors (i.e, uncertainty in the stellar models used for the SFHs) in their results, as well. 
This discrepancy between  the faint-end slopes can be interpreted as a different evolution for the LG dwarfs relative to the field galaxies.

As seen in the middle panel of Figure \ref{fig:lf-parameters}, our characteristic UV magnitude, $M^{*}$, becomes 
brighter, $\Delta M^{*}=-0.7$, at $z=1.9$ relative to $z=1.3$. Also, our characteristic number density, $\phi^{*}$, 
(lower panel of Figure \ref{fig:lf-parameters}) decreases by a factor of 1.5 over this time period. However, both of these 
measurements are dependent on data from other surveys, as our data only sample galaxies fainter than $M^{*}$.

 \begin{deluxetable*}{cccc}{!t}
\tablecaption{UV Luminosity Density\tablenotemark{a}}
\tablewidth{0pt}
\tablehead{\colhead{z} & \colhead{$M<-10$} & \colhead{$M<-17.475(0.04L^{*}_{z=3})$\tablenotemark{b}} & \colhead{$-18.5<M<-12.5$}}\\
\startdata 
$1.0<z<1.6$    &   $1.57\substack{+0.08 \\ -0.08}\times 10^{26}$  & $0.90\substack{+0.06 \\ -0.06}\times 10^{26}$ & $0.90\substack{+0.04 \\ -0.05}\times 10^{26}$   \\
$1.6<z<2.2$    &   $2.84\substack{+0.15 \\ -0.15}\times 10^{26}$  & $1.50\substack{+0.09 \\ -0.09}\times 10^{26}$ & $1.57\substack{+0.08 \\ -0.10}\times10^{26}$   \\
$2.2<z<3.0$   &   $3.13\substack{+0.22 \\ -0.24}\times 10^{26}$  & $1.03\substack{+0.14 \\ -0.19}\times 10^{26}$ & $1.84\substack{+0.13 \\ -0.15}\times 10^{26}$   \\
\enddata 
\tablenotetext{a}{These values are in units of erg s$^{-1}$ Hz$^{-1}$ Mpc$^{-3}$.}
\tablenotetext{b}{For $L^{*}_{z=3}$ we use the measurement from \citet{red09}.}
\label{tab:ldensity}
\end{deluxetable*}

\subsection{UV Luminosity Density}
We use our best LF determinations to derive the comoving UV luminosity density, $\rho_{UV}$, as below:
\begin{equation}
 \rho_{UV}=\int_{L_{\mathrm{faint}}}^{\infty}L\phi(L)\mathrm{d}L=\int_{-\infty}^{M_{\mathrm{faint}}}L(M)\phi(M)\mathrm{d}M
  \label{eq:Ldensity}
 \end{equation}
 where $\phi(L)$($\phi(M)$) is the LF assuming a Schechter function. As an important consequence of the steep faint-end slope 
 of the UV luminosity functions at $1<z<3$, the faint star-forming galaxies have a significant contribution to the total 
 unobscured UV luminosity density at these redshifts. To quantify this, we calculate the cumulative UV luminosity density 
 down to various UV luminosity limits. Figure \ref{fig:lf_all_redshift_nocluster} shows these results for our three redshift ranges. 
 We note that all of these calculations are from our photometric redshift LFs, as they have smaller statistical uncertainties. 
 We normalized our cumulative UV luminosity densities to the corresponding value at $M_{UV}$=-10 assuming 
 that there is no turnover in the LF down to this absolute magnitude. To estimate the 1$\sigma$ uncertainty at each $M_{UV}$, 
 we run a Gibbs sampler (i.e., Markov Chain Monte Carlo sampling) to obtain a sequence of random pairs 
 of ($\alpha$,$M^{*}$) using their 2D joint probability function and then calculate the distribution of UV luminosity density and 
 the corresponding uncertainty. We also incorporate the Poisson uncertainty in quadrature. These 1$\sigma$ uncertainty 
 regions are shaded orange in Figure \ref{fig:lf_all_redshift_nocluster}. The unobscured UV luminosity density measurements 
 are tabulated in Table \ref{tab:ldensity}. To be consistent with previous studies, we also provide the UV luminosity density values integrated down to 0.04$L^{*}_{z=3}$.

The faint dwarf galaxies with UV magnitudes of $-18.5<M_{1500}<-12.5$ covered in this work, comprise the majority of the 
unobscured UV luminosity density at the redshifts of peak star formation activity (58\%, 55\%, and 59\% of the total UV 
luminosity density at $z\sim1.3$, $z\sim1.9$, and $z\sim2.6$, respectively). Therefore, these dwarf galaxies may contribute 
significantly to the total {\it intrinsic} UV luminosity density and thus to the star formation rate density at these epochs. 
However, to quantify this, we need to incorporate the effect of dust reddening and its dependence on galaxy luminosity.

\begin{figure}[!t]
\centering
\includegraphics[angle=0,trim=0cm 0cm 1cm 1cm,clip=true,width=1.\columnwidth]{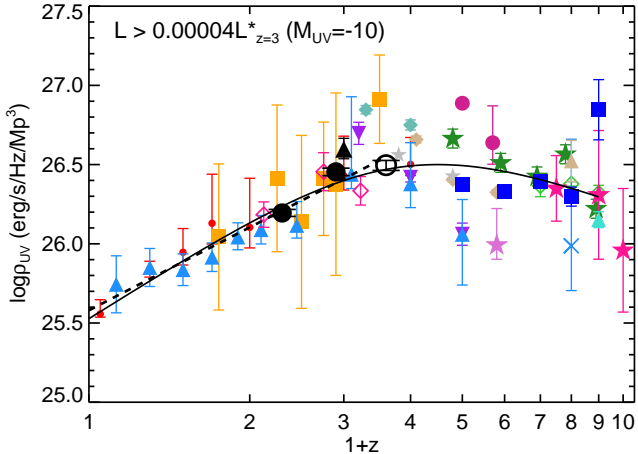}
\centering
\caption{Redshift evolution of the unobscured UV luminosity density measured at rest-frame wavelength of 1500 \AA\ . 
To estimate the $\rho_{UV}$ values, the LFs are integrated down to $M_{UV}=-10$ at all redshifts. The uncertainty 
for each data point is derived by retaining the fractional error of published $\rho_{UV}$ values from each author. 
The black filled circles are derived from our photometric redshift LF estimates. Similar to Figure \ref{fig:lf-parameters}, 
we show our $\rho_{UV}$ measurement at $z=2.6$ with a black open circle as it depends on the choice of $M^{*}$ prior. 
The rest of the symbols are similar to Figure \ref{fig:lf-parameters}, except the red filled circles which are from \citet{sch05} 
using the LF estimates from \citet{arn05,wyd05} (shown with red filled circle and blue asterisk in Figure \ref{fig:lf-parameters}, 
respectively). The dashed line indicates the best-fitting power law to the data points at $z<2.6$. 
The solid line shows the best-fitting function (see Equation \ref{eq:ldensity_evolution}) for the redshift range of $0<z<8$.} 
\label{fig:ld_evolution}
\end{figure}

In order to understand the evolution of the UV luminosity density, we compare our $\rho_{UV}$ measurements with other 
studies at various redshifts. As the value of $\rho_{UV}$ depends on the limiting luminosity, i.e., $L_{faint}$ in 
Equation \ref{eq:Ldensity}, we use the published Schechter parameters from the literature and calculate the UV 
luminosity densities and corresponding uncertainties by integrating down to the same $M_{\mathrm{faint}}=-10$. 
We should note that there is no straightforward way to estimate the $\rho_{UV}$ uncertainties as necessary 
information for these measurements such as covariance between Schechter parameters are not usually provided in 
the literature. But to assign uncertainty to each $\rho_{UV}$ measurement in a consistent way, we use the same 
methodology as \citet{mad14}. We assume that the fractional error, i.e., $\Delta\rho_{UV}/\rho_{UV}$, provided by 
each author is fixed and thus derive the corresponding uncertainty for our $\rho_{UV}$ value with $M_{faint}=-10$. 
Figure \ref{fig:ld_evolution} illustrates these measurements. As seen in many previous studies \citep[e.g., ][]{cuc12}, 
the unobscured (i.e, uncorrected for dust extinction) UV luminosity density rises from $z=0$ to $z=2.0$ where it 
reaches its peak and starts to decline after $z=3$ \citep[e.g.,][]{oes10a,fin15}. As shown in Figure \ref{fig:ld_evolution}, 
our $\rho_{UV}$ points (black filled circles) follow the similar trend as seen by previous determinations. 
However, our measurements show a more rapid evolution from $z=1.3$ to $z=1.9$ followed by a slower evolution up to $z=2.6$. 

We emphasize that the unobscured $\rho_{UV}$ evolution rate and the exact location of the peak depends on the 
wavelength \citep{tre12} where $\rho_{UV}$ is being measured, and the limiting luminosity, i.e., $L_{faint}$ in 
Equation \ref{eq:Ldensity}. Therefore, to find the best-fitting function describing the evolution of unobscured 
UV luminosity density between $z=0-2.6$, we only include the results from the literature at the same 
wavelength (1500\AA\/) and integrated down to the same magnitude of $M_{UV}=-10$ through our own compilation. 
Fitting a power law, we find $\rho_{UV}=25.58\times(1+z)^{1.74}$ incorporating the data points 
from \citet{sch05}(red filled circle); \citet{dah07} (pink open diamond); \citet{oes10a} (for photometric redshift sample, orange filled square) and \citet{cuc12} (blue filled triangle).

 In addition, to study the evolution of $\rho_{UV}$ for the whole redshift range from $z=0$ to $z=8$, we fit a 
 function used by \citet{mad14} as shown below. For the higher redshifts, we incorporate the data points 
 from \citet{mcl09,mcl13}(green open diamond), \citet{bou15}(green filled star) and \citet{par16}(gray filled star), as well as the data points that we used for the power law.
\begin{equation}
 \rho_{UV}(z)=a\frac{(1+z)^{b}}{1+[(1+z)/d]^{c}} \mathrm{erg\ s^{-1}\ Hz^{-1}\ Mpc^{-3}}
  \label{eq:ldensity_evolution}
 \end{equation}
where we derive $a=0.34\pm0.04$, $b=2.14\pm0.27$, $c=3.41\pm0.23$ and $d=3.86\pm0.63$. We emphasize 
that these best-fit values describe the $\rho_{UV}$ evolution assuming a limiting magnitude of $M_{UV}=-10$, 
dramatically fainter than typical limits used in previous studies \citep[$\sim$-17.5,][]{mad14}. Because we do 
not account for an increase in the uncertainty of $\rho_{UV}$ at low luminosities, we add in quadrature $12\%$ 
uncertainty to all of the data points to keep the reduced chi-squared equal to one.

\subsection{No Turnover in the UV LF}
\label{subsec:turnover}
Our observations have now reached the very faint luminosities where some simulations predict a turnover in the UV LF. 
Though our steep LFs extend down to $M_{UV}=-12.5$ and we showed that the faint bins with large completeness corrections are not affecting 
the faint-end slope fit (see \ref{subsec:mle}), they may affect our interpretation of whether or not there is a turnover. 
As see in Figure \ref{fig:lf}, we can rule out the possibility of a turnover in the LF at magnitudes brighter than $M_{UV}<-14$, 
because one would need an unphysical large systematic effect to cause a turnover at this magnitude.
This conclusion is in conflict with the results of 
recent cosmological hydrodynamical simulation by \citet{kuh13}, who predict a turnover at $M_{UV}=-16$ in the $z\sim2.5$ 
UV LF. Implementing an H$_{2}$-regulated star formation model, they predict that the star formation is suppressed 
in dwarf galaxies ($M_{UV}>-16$), because their gas surface density is below what is required to build a substantial 
molecular fraction. A similar tension between the observed UV LF and the turnover predicted by recent theoretical results has also been seen at higher redshifts \citep[e.g.,][]{jac13,liv16}.

The presence of a turnover in the UV LF might also be used to constrain warm dark matter (WDM) models. \citet{men16} 
provide a limit on the WDM particle mass by comparing the WDM halo mass function and the number density of ultra-faint 
galaxies derived from the UV LF in A14. The constraints can now be significantly improved given the much larger sample in this survey. 

\section{Conclusion}
\label{sec:conclusion}
We have obtained deep near-UV imaging of three lensing clusters, two from the HFF surveys (A2744 and MACSJ0717)
 and A1689, to study the evolution of the UV LF during the peak epoch of cosmic star formation at $1<z<3$. 
 Combining deep data with strong gravitational lensing magnification, we obtain a large sample ($780$) of ultra-faint 
 galaxies with $M_{UV}<-12.5$ at $1<z<3$, using the photometric redshift selection. We perform an extensive set 
 of simulations to compute the completeness correction required for the LF measurements. We summarize our conclusions below:
\begin{itemize}
\item We derive the best Schechter fit to each UV LF using a maximum likelihood technique considering various sources 
of uncertainty including the lensing models. Thanks to the lensing magnification, we can extend the UV LF 
measurements down to very faint luminosities of $M_{UV}=-12.5$ at $1<z<3$. Consequently, we find a robust estimate 
of the UV LF faint-end slope to be $\alpha=-1.56\pm0.04$, $\alpha=-1.72\pm0.04$ and $\alpha=-1.94\pm0.06$ 
at $1.0<z<1.6$, $1.6<z<2.2$ and $2.2<z<3.0$, respectively. Our $\alpha$ measurements at $z\sim1.3$ and $z\sim1.9$ 
are consistent with previous studies of \citet{red09,oes10a}. But for $z\sim2.3$, we have a steeper faint-end slope 
than previous studies. Our determinations of the UV LFs show a rapid evolution in the faint-end slope toward steeper 
values from $z=1.3$ to $z=2.6$. 
In addition, when we run a test to minimize the systematic effects by excluding galaxies and volumes $<50\%$ completeness, we still derive 
steep faint-end slopes of $\alpha=-1.55\pm0.06$, $-1.69\pm0.07$ and $-1.79\pm0.08$ at $z\sim1.3$, $1.9$ and $2.6$, respectively.
However, for a better determination of the LF parameters, we need a better understanding of the size and color distribution of these faint galaxies.

\item To understand the effect of different selection techniques on the UV LF, we use a two color ``dropout" selection of 
Lyman break galaxies at redshifts similar to our photometric redshift samples. After correcting for incompleteness 
and then finding the best fit Schechter parameters, our LBG and photometric redshift LFs are in $2\sigma$ agreement. 

\item We integrate our UV LFs down to a magnitude limit of $M_{UV}=-10$ and find the UV luminosity density to 
be $\rho_{UV}=1.57\substack{+0.08 \\ -0.08}\times 10^{26}$, $2.84\substack{+0.15\\ -0.15}\times 10^{26}$ and 
$3.13\substack{+0.22 \\ -0.24}\times 10^{26}$ erg s$^{-1}$ Hz$^{-1}$ Mpc$^{-3}$ at $z\sim1.3$, $z\sim1.9$ and $z\sim2.6$, respectively. We show that the faint star-forming galaxies 
with $-18.5<M_{UV}<-12.5$, contribute the majority of the total unobscured UV luminosity density during the peak epoch of cosmic star formation. 

\item Though some models of warm dark matter and some prescriptions for H$_{2}$-regulated star-formation 
predict a turnover in the UV LF, we see no evidence for a turnover down to $M_{UV}=-14$ at $1<z<3$.  

\end{itemize}

This study highlights the power of gravitational lensing to produce a robust constraint on the faint-end of the LF. 
However, as mentioned in Section \ref{sec:photz_vs_lbg}, this analysis still suffers from uncertainties that are 
systematic, rather than statistical. To overcome these uncertainties, in the future, we require precise measurements 
of size distribution and dust reddening at low luminosities.\

\
We thank the referee for a careful reading and useful comments that improved this paper.
The authors are grateful to the STScI and HFF team for obtaining and reducing the {\it HST} images. A.A. would like to thank 
Pascal Oesch for providing their individual measurements for photometric redshift LFs, Takatoshi Shibuya for 
sending size measurements, Jose Diego for providing us a list of A1689 multiple images, as well as John Blakeslee and Karla Kalamo for providing us a list of globular clusters of A1698. 
A.A. also thanks Daniel Weisz for his valuable comments as well as Naveen Reddy, 
Nader Shakibay Senobari, Mario De Leo, Ali Ahmad Khostovan and Kaveh Vasei for useful conversations. MJ acknowledges support from 
the Science and Technology Facilities Council [grant number ST/L00075X/1 \& ST/F001166/1]. ML acknowledges the 
 Centre National de la Recherche Scientifique (CNRS) for its support. This work is based on observations with the NASA/ESA Hubble Space Telescope, obtained at the Space Telescope Science Institute, which is operated by the Association of Universities for Research in Astronomy, Inc., under NASA contract NAS 5-26555.


\appendix
\section{A. Lyman break Selection}
\label{appendixA}
\begin{figure*}[!t]
\centering
\includegraphics[angle=0,trim=0cm -1cm 0cm 0cm,clip=true,width=1.0\columnwidth]{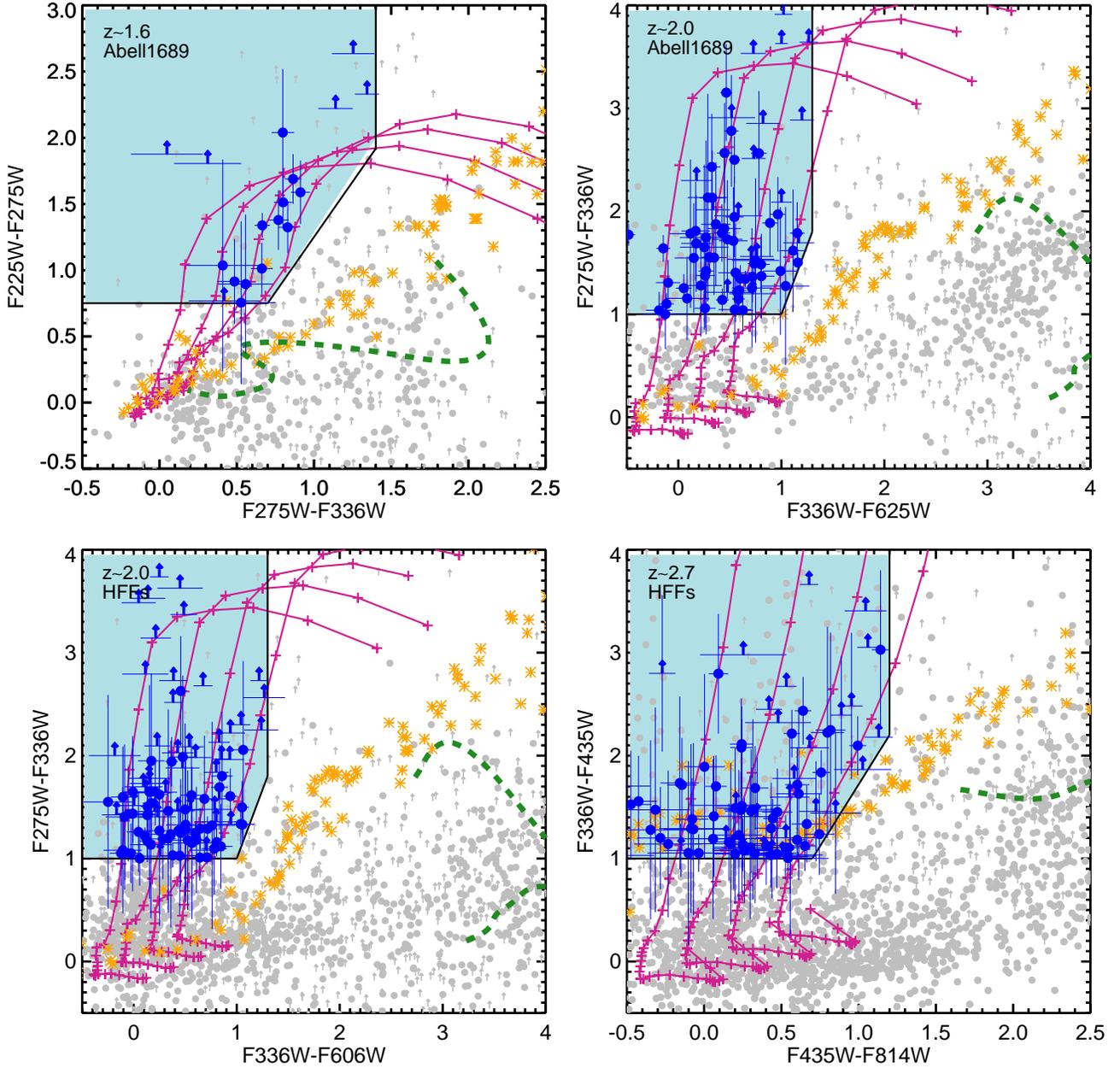}
\centering
\vskip-0.4in
\caption{Color-color selection of Lyman break galaxies for F225W- (left), F275W- (middle) and F336W-dropouts (right). A1689 and HFF LBG selections are shown separately in the upper and lower panels, respectively. Gray circles are all of the objects in the corresponding catalogs. The orange asterisks are stars from \citet{pic98}. The green dashed line shows the color track of low redshift ($0<z<1$) elliptical galaxies from \citet{col80}. The violet lines are star-forming tracks with different dust reddening values of E(B-V)=[0,0.1,0.2,0.3]. The blue region shows the selection criteria. The blue symbols are the LBG candidates with 5$\sigma$ detection in two bands redward of Lyman limit (see text). The blue arrows represent 1$\sigma$ lower limits.} 
\label{fig:lbg}
\end{figure*}

In this section, we outline our selection criteria to find Lyman break galaxies \citep{ste99}. We adopted a standard color-color diagram to sample the UV continuum break in the SED of high redshift galaxies. As shown in Figure \ref{fig:lbg}, the selection region is defined based on the location of tracks of star forming  galaxies in the color-color plot. The star-forming tracks are predictions from \citep{bru03} synthetic stellar population models assuming a constant star formation history, $0.2\ Z_{\sun}$ and an age of 100 Myr with different color excess of E(B-V)=[0,0.1,0.2,0.3]. 
In the next subsection, we summarize the selection criteria we use to identify the z $\sim 1-3$ galaxies. As A1689 is observed with different sets of filters than the HFFs, it is not possible to use the same color criteria for all of these fields. Therefore, we construct analogous selection criteria as below.

For F225W dropout sources, considering that A1689 is the only field where F225W images are available, the selection criteria are as below, 
\begin{displaymath}
F225W-F275W > 0.75
\end{displaymath}
\begin{displaymath}
F275W-F336W < 1.4
\end{displaymath}
\begin{displaymath}
F225W-F275W > 1.67 (F275W-F336W) - 0.42
\end{displaymath}
\begin{displaymath}
S/N(F275W) > 5 \hspace{0.5cm} , \hspace{0.5cm} S/N(F336W) > 5
\end{displaymath}
\begin{equation}
\label{equ:color-selection1}
\end{equation}
These color criteria, which are identical to \citet{oes10a} find 31 galaxy candidates in A1689.

For F275W dropout sources, the selection criteria for A1689 with F625W-band imaging are 
\begin{displaymath}
F275W-F336W > 1
\end{displaymath}
\begin{displaymath}
F336W-F625W < 1.3
\end{displaymath}
\begin{displaymath}
F275W-F336W > 2.67 (F336W-F625W) - 1.67
\end{displaymath}
\begin{displaymath}
S/N(F336W) > 5 \hspace{0.5cm} , \hspace{0.5cm} S/N(F625W) > 5
\end{displaymath}
\begin{equation}
\label{equ:color-selection2}
\end{equation}
 These color criteria, which are identical to what we used before in A14 find 99 galaxy candidates in A1689. For F275W dropout sources, the selection criteria for HFFs where F606W-band imaging are available instead of F625W-band data, we use identical selection criteria as for A1689. In total, these color criteria find 230 galaxy candidate over three clusters. 99 of these candidates are from A1689 in comparison with 84 candidates in A14, because we added 14 orbits to the 4 orbits of F336W image that we used in A14.

 For F336W dropout sources, the selection criteria for HFFs are
 \begin{displaymath}
F336W-F435W > 1
\end{displaymath}
\begin{displaymath}
F435W-814W < 1.2
\end{displaymath}
\begin{displaymath}
F336W-F435W > 2.4 (F435W-F814W) - 0.68
\end{displaymath}
\begin{displaymath}
S/N(F435W) > 5 \hspace{0.5cm} , \hspace{0.5cm} S/N(F814W) > 5
\end{displaymath}
\begin{equation}
\label{equ:color-selection3}
\end{equation}
These criteria find 189 galaxy candidates over HFFs. Similar to our $z \sim 2.6$ photometric redshift sample (see Section \ref{sec:selection}), we do not include A1689 in our F336W-dropout sample as there is contamination from cluster members. As discussed in Section \ref{sec:selection}, we remove all of the multiple images corresponding to a single source except the brightest image. We then use the same identification scheme as our photometric redshift samples to remove contamination from stars, stellar spikes and spurious detections from low-redshift bright galaxies. As seen in Figure \ref{fig:lbg}, the stellar sequence (orange asterisks) enters the selection region of F225W- and F336W-dropouts, resulting a contamination of $3.2\%$ and $1.6\%$ of stars, respectively. We also excluded a low contamination of $1.3\%$ and $4.2\%$ from the stellar spikes and spurious objects in the F275W- and F336W-dropouts, respectively. 

\begin{figure*}[!t]
\centering
\includegraphics[angle=0,trim=0cm -1cm 0cm 0cm,clip=true,width=0.65\columnwidth]{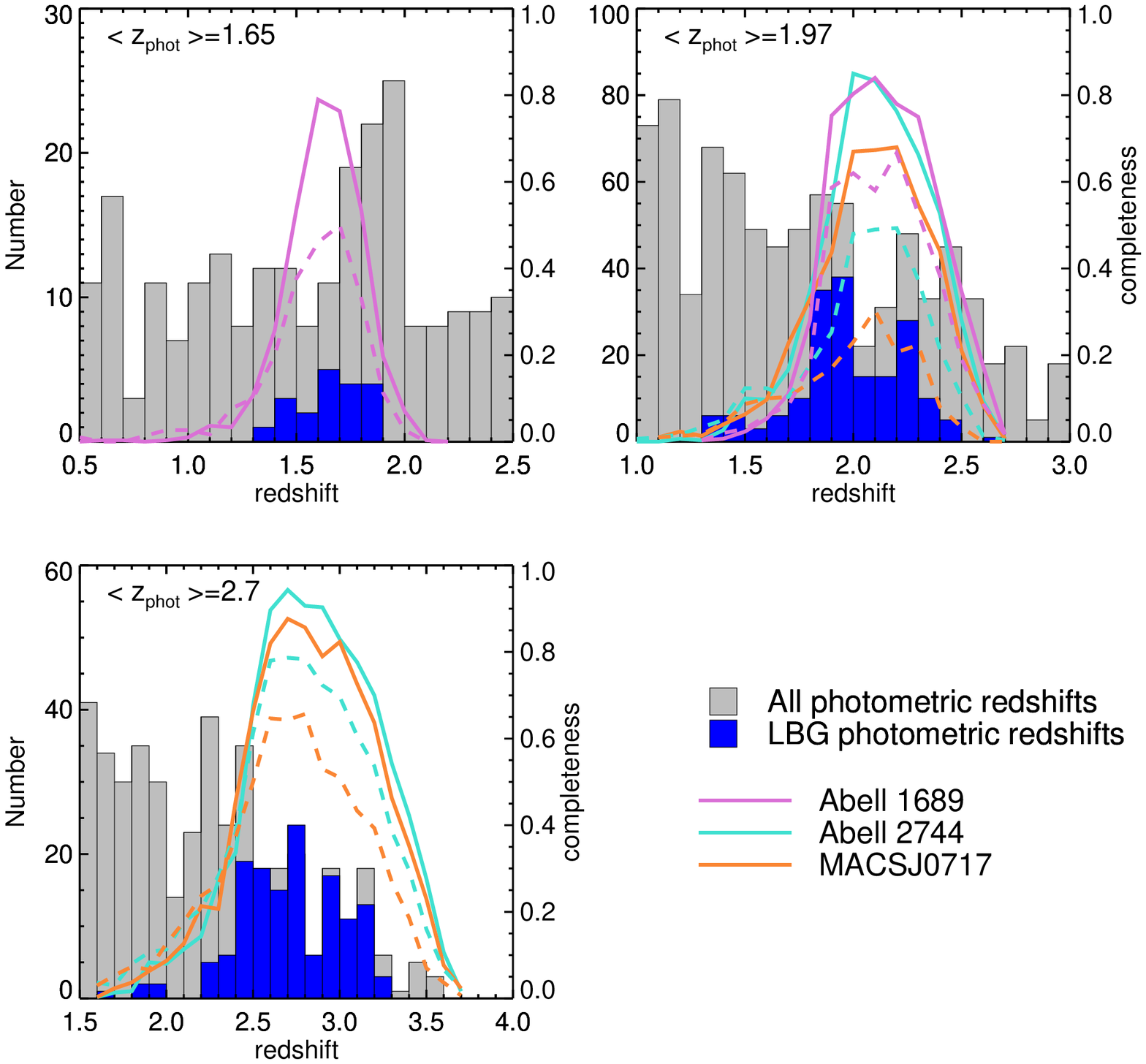}
\centering
\vskip-0.4in
\caption{Photometric redshift distribution of LBGs. The gray histogram shows the photometric redshift distribution of all objects in the fields. The blue histogram highlights the photometric redshift distribution of those galaxies selected as LBG. The solid and dashed lines show the simulated completeness distribution over redshift for a galaxy with $\mu_{\mathrm{mag}}=1.0$ and $m_{UV}=27$ and 28, respectively. The purple, blue and orange colors present the completeness distribution for A1689, A2744 and MACSJ0717, respectively. The right-hand axis shows the completeness values.} 
\label{fig:z_distribution}
\end{figure*}

In addition, our photometric redshift measurements show that the fraction of low-redshift interlopers in the LBG samples are low. We find that only $9.7\%$, $7.8\%$ and $5.8\%$ of our F225W-, F275W- and F336W-dropout samples are low-redshift interloper with $z<1.0$, $z<1.3$ and $z<1.5$, respectively. As illustrated in Figure \ref{fig:z_distribution} and explained in next section, we derive these redshift cuts using the expected redshift distribution from our completeness simulations for dropout samples.
 
\section{B. Completeness Simulation for LBG LF}
\label{lbg_comp}
Following the analysis for our photometric redshift sample, we run the same Monte Carlo simulation to assess the completeness values, C(m, z, $\mu_{\mathrm{mag}}$), for the Lyman break samples. As described in details in Section \ref{sec:completeness}, after generating the random artificial galaxies with similar properties as observed sources, we require the same selection criteria (Equations A1 to A3) as we used for the observed LBGs. Figure \ref{fig:z_distribution} illustrates the completeness distribution for two intrinsic apparent magnitudes (i.e., before magnification) of $m_{UV}=27, 28$ and magnification $\mu_{\mathrm{mag}}=1.0$ mag. To compare with the observed galaxies, we over-plot the photometric redshift distribution of the whole catalog together with the subsample selected as LBG. As seen in this figure, the redshift distribution of LBGs (blue histograms) relative to the redshift distribution of all galaxies in the field (grey histograms) are consistent with the completeness calculations.

\section{C. New Multiply Imaged Systems}
\label{appendixC}
\begin{figure*}[!t]
\centering
\includegraphics[angle=0,trim=0cm 0cm 0cm 0cm,clip=true,width=1.0\columnwidth]{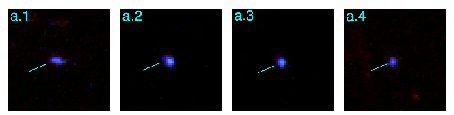}
\includegraphics[angle=0,trim=0cm 0cm 0cm 0cm,clip=true,width=1.0\columnwidth]{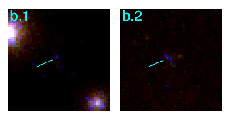}
\includegraphics[angle=0,trim=0cm 0cm 0cm 0cm,clip=true,width=1.0\columnwidth]{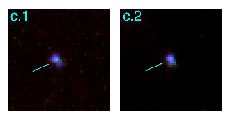}
\includegraphics[angle=0,trim=0cm 0cm 0cm 0cm,clip=true,width=1.0\columnwidth]{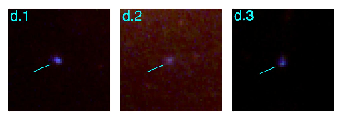}
\includegraphics[angle=0,trim=0cm 0cm 0cm 0cm,clip=true,width=1.0\columnwidth]{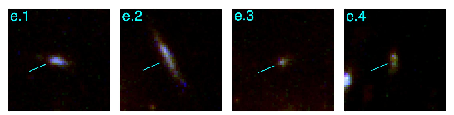}
\caption{Newly-identified multiply imaged systems in A1689. The arrow in each stamp denotes the position of each multiple image. The color image is a combination of F475W(blue), F625W(green) and F814W(red) filters. We note that the reddening in d.2 image is due to a nearby cluster member. The size of each cutout is $2\arcsec$.} 
\label{fig:stamp}
\end{figure*}

\begin{figure*}[!t]
\centering
\includegraphics[angle=0,trim=0cm 0cm 0cm 0cm,clip=true,width=1.0\columnwidth]{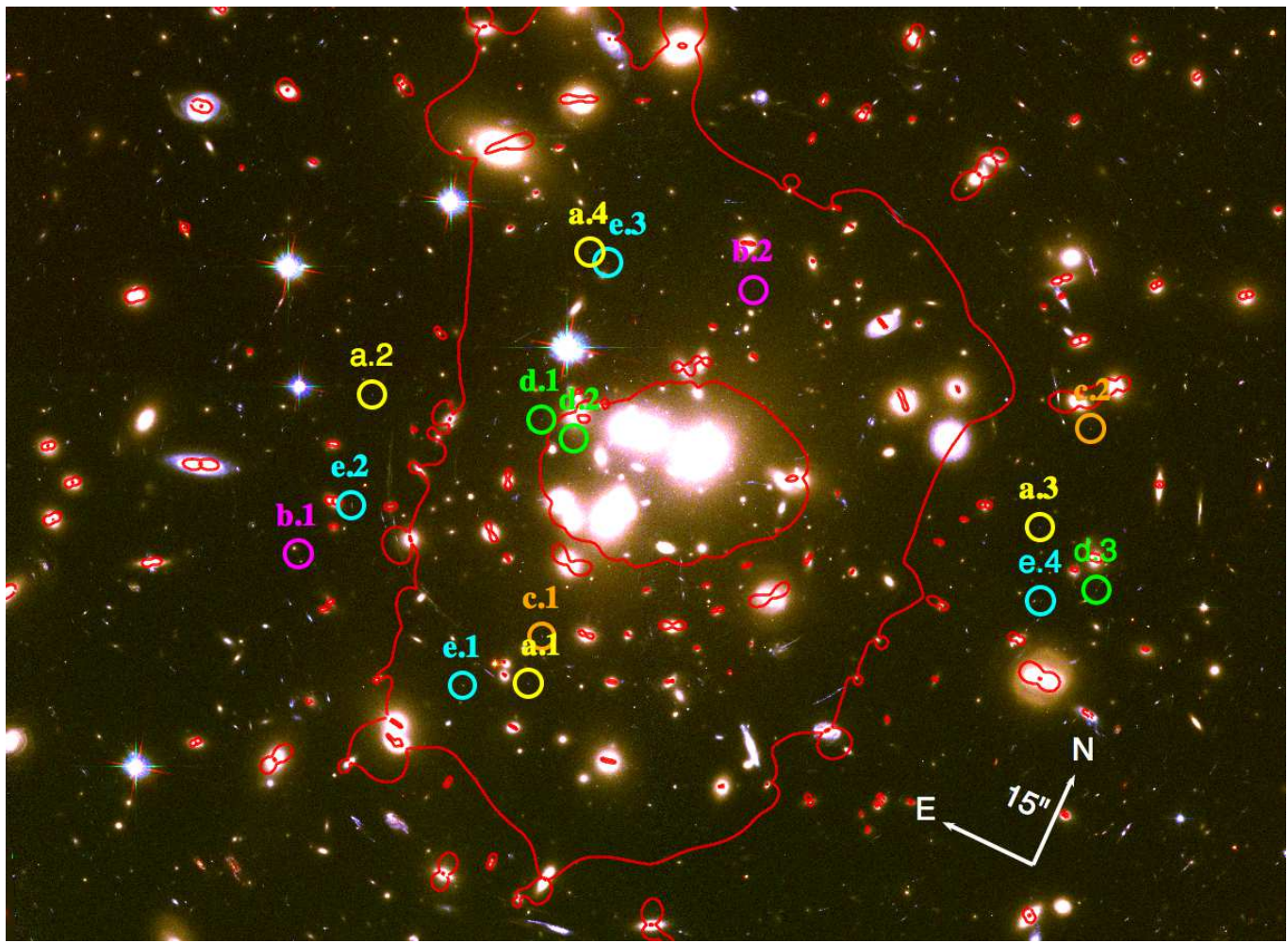}
\caption{The color image is a composite of F475W(blue), F625W(green) and F814W(red) filters. The contours show the critical lines for sources at z=2.5. The circles denote the positions of the newly found multiple images. A compass provides the orientation and the lengths of the arrows show the $15\arcsec$ scale. Some of the labels have been offset slightly.} 
\label{fig:critical}
\end{figure*}

As described in Section \ref{subsec:multi}, we identify 5 new multiply lensed system candidates in A1689. Table \ref{tab:multi}  summarizes these systems where we provide their photometric redshift estimates and color measurements, as well. Because one of the primary indicators of multiple images is their uniform colors (i.e., magnification is independent of wavelength), we show their RGB composite image (see Figure \ref{fig:stamp}) combining F814W, F625W and F475W data as red, green and blue filters. In Figure \ref{fig:critical}, we show the positions of all 5 new systems on a color-composite image of A1689. We also overplot the critical lines at $z=2.5$. The RGB colors are similar to Figure \ref{fig:stamp}.   

\begin{deluxetable*}{ccccccccc}
\tablecaption{New Multiply Imaged Systems in A1689}
\tablewidth{0pt}
\tablehead{\colhead{system} & \colhead{R.A.(J2000)} & \colhead{Dec(J2000)} & \colhead{$z_{phot}$} & \colhead{$\Delta z_{phot}$} &
	\colhead{F336W-F475W} & \colhead{F475W-F625W} & \colhead{F625W-F775W} & \colhead{F775W-F814W}}
\startdata a.1 & 197.875427 & -1.353059 & 2.33 & 0.13 & 0.72$\pm$0.12 & -0.19$\pm$0.12 & -0.05$\pm$0.15 & -0.01$\pm$0.12 \\
a.2 & 197.886612 & -1.344707 & 1.90 & 0.07 & 0.30$\pm$0.09 & -0.09$\pm$0.10 & -0.08$\pm$0.12 & 0.03$\pm$0.10  \\
a.3 & 197.858505 & -1.337834 &  2.33 & 0.10 & 0.67$\pm$0.11 & -0.11$\pm$0.11 & -0.12$\pm$0.13 & -0.02$\pm$0.11  \\
a.4 & 197.880814 & -1.335316 &  2.20 & 0.17 & 0.65$\pm$0.15 & -0.08$\pm$0.14 & -0.01$\pm$0.17 & -0.13$\pm$0.14 \\
b.1 & 197.886612 & -1.352190 &  1.87  & 0.27 & 0.48$\pm$0.29 & -0.21$\pm$0.33 & -0.06$\pm$0.41 & -0.08$\pm$0.33 \\
b.2 & 197.873810 & -1.333846 &  2.02  & 0.10 &  -0.03$\pm$0.18 & -0.36$\pm$0.28 & 0.06$\pm$0.34 & -0.13$\pm$0.27\\
c.1 & 197.875732 & -1.350983 &  2.47  & 0.10 & 1.56$\pm$0.22 &  0.07$\pm$0.11 & 0.02$\pm$0.12 & -0.07$\pm$0.10 \\
c.2 & 197.858337 & -1.333123 &  2.40 & 0.06 &  1.30$\pm$0.17 & 0.23$\pm$0.10 & 0.17$\pm$0.10 & -0.21$\pm$0.08 \\
d.1 & 197.879654 & -1.342635 &  2.61 & 0.10 & 1.92$\pm$0.30 & 0.01$\pm$0.12 & -0.16$\pm$0.14 & 0.01$\pm$0.12\\
d.2 & 197.878067 & -1.342786 &  2.57  & 0.19 & 2.08$\pm$0.54 & 0.02$\pm$0.17 & -0.02$\pm$0.19 & 0.15$\pm$0.15 \\
d.3 & 197.855209 & -1.339240 &  2.30  & 0.09  & 0.62$\pm$0.12 & 0.11$\pm$0.11 & -0.04$\pm$0.12 & -0.19$\pm$0.10 \\
e.1 & 197.877899 & -1.354296 &  3.19  & 0.19 & $>$3.32 &  0.67$\pm$0.09 & 0.10$\pm$0.08 & -0.07$\pm$0.07\\
e.2 & 197.885437 & -1.349361 &  3.23 & 0.18 & $>$3.65  &   0.64$\pm$0.08 &  0.05$\pm$0.07 & 0.05$\pm$0.06 \\
e.3 & 197.879929 & -1.335399 &  3.11 & 0.26 &  $>$2.39  &   0.76$\pm$0.17 & 0.13$\pm$0.13 & 0.04$\pm$0.10\\
e.4 & 197.857147 & -1.340662 &  3.32 & 0.23 & $>$2.86  -&  0.89$\pm$0.13 & 0.02$\pm$0.10 & 0.08$\pm$0.08   

\enddata
\label{tab:multi}
\end{deluxetable*}

\end{document}